# Eroding dipoles and vorticity growth for Euler flows in $\mathbb{R}^3$: The hairpin geometry as a model for finite-time blowup

**Stephen Childress**[1] ‡ **and Andrew D. Gilbert**[2] §

[1]Courant Institute of Mathematical Sciences,
New York University, New York, NY 10012, USA
[2]Department of Mathematics,
College of Engineering, Mathematics, and Physical Sciences,
University of Exeter, Exeter EX4 4QF, UK

‡ childress@cims.nyu.edu
§ A.D.Gilbert@exeter.ac.uk

**Abstract.**  A theory of an eroding "hairpin" vortex dipole structure in three dimensions is developed, extending our previous study of an axisymmetric eroding dipole without swirl. The axisymmetric toroidal dipole was found to lead to maximal growth of vorticity, as $t^{4/3}$. The hairpin is here similarly proposed as a model to produce large "self-stretching" of vorticity, with the possibility of finite-time blow-up. We derive a system of partial differential equations of "generalized" form, involving contour averaging of a locally two-dimensional Euler flow. We do not attempt here to solve the system exactly, but point out that non-existence of physically acceptable solutions would most probably be a result of the axial flow. Because of the axial flow the vorticity distribution within the dipole eddies is no longer of the simple Sadovskii type (vorticity constant over a cross-section) obtained in the axisymmetric problem. Thus the solution of the system depends upon the existence of a larger class of propagating two-dimensional dipoles.

The hairpin model is obtained by formal asymptotic analysis. As in the axisymmetric problem a local transformation to "shrinking" coordinates is introduced, but now in a self-similar form appropriate to the study of a possible finite-time singularity. We discuss some properties of the model, including a study of the helicity and a first step in iterating toward a solution from the Sadovskii structure. We also present examples of two-dimensional propagating dipoles not previously studied, which have a vorticity profile consistent with our model. Although no rigorous results can be given, and analysis of the system is only partial, the formal calculations are consistent with the possibility of a finite time blowup of vorticity at a point of vanishing circulation of the dipole eddies, but depending upon the existence of the necessary two-dimensional propagating dipole. Our results also suggest that conservation of kinetic energy as realized in the eroding hairpin excludes a finite time blowup for the corresponding Navier-Stokes model.

# 1. The eroding hairpin dipole

## 1.1. Introductory comments

In our earlier paper, [13] (hereafter denoted as I), the development of an eroding vortex dipole was studied with the assumption of axisymmetric flow without swirl (AFWOS). We described the "snail" geometry of the dipole, noted the relation to the Sadovskii vortex dipole, and calculated the growth of vorticity as $t^{4/3}$. We also found that the circulation remaining in each half of the "head" of the snail fell as $t^{-2/3}$. These calculations are a model for the ultimate fate of two equal and opposite ring vortices which collide while maintaining axial symmetry.

The studies in [10] and I can be summarized as consisting of the following steps:

- Assume AFWOS governed by the Euler equations.

- Optimize the rate of growth of vorticity in a mirror-symmetric toroidal dipole.

- Impose conservation of kinetic energy through erosion of the dipole.

- Identify the asymptotic structure of the dipole for large time as a perturbed Sadovskii vortex (the snail).

- Verify the asymptotic results by numerical experiment.

We show in figure 1(a) a representation of the eroding toroidal dipole.

In the present paper, our aim is to extend this analysis to a more general three-dimensional geometry. We shall retain the basic mirror symmetry of the dipole with respect to its plane of motion. However the center of the dipole cross-section is no longer constrained to lie on a circle in the plane of motion, see figure 1(b). We shall see that the generalization leads to the possibility of much more rapid local growth of vorticity, and possibly a finite time singularity. If a singularity occurs in the hairpin, it does so at a single point where vorticity is infinite but circulation in each constituent head is zero. This is an unusual and unexpected feature of the proposed singularity.

Be forewarned that much of the notation for the three-dimensional (3D) problem is new, and is not to be confused with that used in the axisymmetric case as treated in I. The geometry will be more general and the asymptotics are now for $t \to 0-$, the proposed singularity time, rather than $t \to \infty$.

A number of important new features of the developing dipole must now be considered. In the first place, there is the issue of local two-dimensionality of the flow. As colliding vortices form a spreading axisymmetric dipole, the product of the curvature $1/R$ of the toroidal dipole, times the radius of the cross-section of the eroding dipole, decreases as $R^{-7/4}$, so local two-dimensionality may be said to be asymptotic in time. It will be shown that local two-dimensionality may in fact be maintained more generally, allowing an almost two-dimensional dynamical model. With core deformation and local two-dimensionality under control, there is hope for a direct analysis of breakdown in three dimensions.

There is a second important issue which arises in the 3D problem, and that is the *axial flow* within the eddies, which occurs in the dipole owing to the development of an axial pressure gradient. This flow can stretch (or contract) vortex lines and thereby modify the dipole away from the Sadovskii form. This in turn can change how the structure moves, possibly changing its defining curve and perhaps expelling a singularity. We shall see that this leads to a difficult problem of generalized analysis of vortical structure close to regions of closed streamlines. The object of the present paper is to develop analysis sufficient to illustrate these difficulties and then to provide models suggestive of the effect of axial flow. A third issue, related to the existence of axial flow, is the fact that the loss of vorticity volume associated with local conservation of kinetic energy may now involve axial flux.

Given this general setting, we shall present an approximate asymptotic analysis of blowup. There is no claim here to rigor, but we do seek to develop enough analysis to uncover some positive aspects of our proposal, as well as to point out some possible pitfalls. We shall study the limiting case of $\tau \equiv -t \to 0+$ within the context of a similitude in which all variables are scaled by fractional powers of $\tau$. In the analysis of singular structures with this similitude, there invariably arises the possibility of variation on a secondary time scale $\bar{\tau} \equiv -\log \tau$. We shall disallow such variation and in so doing restrict the discussion to what might be called "steady" singular behaviour.

In spite of these important differences the present study will nevertheless follow in many respects the structure of I, consisting of the following steps:

- Apply the similitude appropriate to the local conservation of kinetic energy of a tubular dipole placed on the center curve (sections 1.2, 1.3). The similarity variables are chosen to explicitly allow a finite-time singularity. Exhibit the shape of the hairpin.

- Localize Euler's equations in a neighborhood of the hairpin and verify the structure as locally two-dimensional. Study the similitude of the "snail" geometry of the eroding dipole as a perturbed Sadovskii vortex (sections 2.1, 2.2).

- Exhibit Euler's equations with the hairpin similitude in appropriate curvilinear coordinates. Introduce an asymptotic expansion near the singularity time and the derivation of dynamical equations as compatibility conditions under contour averaging (section 2.3).

- Establish blow-up under the condition that there is no axial flow in the dipole, and outline the asymptotic potential flow exterior to the hairpin (sections 2.4, 2.5).

- Carry out the derivation of compatibility equations by contour averaging, introduce shrinking variables to account for erosion, and specialize the system to the case of "steady" blow-up (section 3).

- Carry out an approximate study of the possible effect of axial flow on the structure of the snail (section 4).

- Extend the Sadovskii dipole to a one-parameter family of dipoles with non-constant vorticity in each lobe. We shall find that the branch of solutions so obtained cannot be extended to reach the vorticity structure produced by the axial flow, at least in the limited calculations carried out here (section 5).

Finally, section 6 discusses the results and the limitations of our approach, and assesses the implications of our paper for the problem of Euler blowup.

### 1.2. The center curve

The *center curve* $\mathscr{C}(t)$, used to determine the position of a locally 2D vortex structure, was introduced by [10] (see also [11]). This is a curve confined to the $(x, z)$ plane, depicted in figure 1(c). The locally 2D vortex dipole, symmetric with respect to that plane (as in figure 1(a)), is aligned with the curve. From the dynamics of the dipole we deduce a law of motion for $\mathscr{C}(t)$ of the form

$$\frac{\partial \mathbf{x}}{\partial t}\Big|_{s_0} = U\mathbf{n} + W\mathbf{t}, \quad U < 0, \tag{1}$$

where $s_0$ is a Lagrangian point on the curve and $\mathbf{t}, \mathbf{n}$ are the unit tangential and normal vectors there. $U, W$ are to be regarded as functions of $s$ or $s_0$ and $t$. Note that we have retained the conventional definition of $\mathbf{n}$, which forces us to deal with a negative $U$.

Using (1) and its time derivative, together with the Frenet–Serret equations, we obtain the following system determining the evolution of $\mathscr{C}(t)$:

$$\frac{\partial J}{\partial t}\Big|_{s_0} = \Big(\frac{\partial W}{\partial s} - U\kappa\Big)J, \tag{2}$$

$$\frac{\partial \kappa}{\partial t}\Big|_{s_0} - W\frac{\partial \kappa}{\partial s} - U\kappa^2 - \frac{\partial^2 U}{\partial s^2} = 0, \tag{3}$$

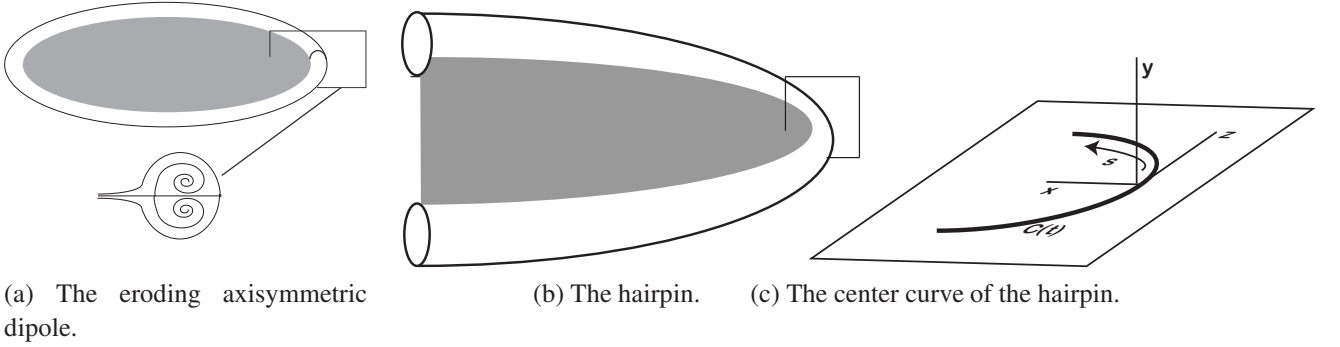

(a) The eroding axisymmetric dipole.

(b) The hairpin.

(c) The center curve of the hairpin.

Figure 1: (a) The axisymmetric dipole is shown with the tail in gray fill. The "snail" geometry shown here is the instantaneous streamline pattern relative to "shrinking" coordinates. (b) The hairpin consists of a similar dipole with the hairpin similitude placed along the center curve (c).

where $\kappa$ is the line curvature, $s$ is arc length, and $J = \partial s / \partial s_0$ is the Jacobian.

To derive these we first take the derivative of (1) with respect to $s_0$ to obtain

$$\frac{\partial (J\mathbf{t})}{\partial t}\Big|_{s_0} = \Big(\frac{\partial W}{\partial s} - JU\kappa\Big)\mathbf{t} + \Big(J\frac{\partial U}{\partial s} + W\kappa J\Big)\mathbf{n}. \tag{4}$$

The tangential component of this equation yields (2). Next, take the normal component of the derivative of (4) with respect to $s_0$. This yields,

$$\frac{\partial (J\kappa)}{\partial t}\Big|_{s_0} = J\frac{\partial^2 U}{\partial s^2} + \frac{\partial W}{\partial s}J\kappa + WJ\frac{\partial \kappa}{\partial s}. \tag{5}$$

Use of (2) in (5) gives (3).

### 1.3. Similitude and the basic hairpin

We summarize here the general results described in [10] involving the parameter $\beta$, specialized to the case $\beta = 4$. In I we showed that the geometry of the axisymmetric eroding dipole ring of radius $R(t)$ was fixed by the requirement of local conservation of kinetic energy plus the relation between local stretching and local axial vorticity. If the lateral dimension of the dipole cross section $\sim a$, then the relations $\omega \sim J$, $U \sim \omega a$, $JU^2a^2 \sim 1$ follow, implying $J \sim U^4$. Balancing the terms of (2),(3), for a hairpin singular at $t = 0-$ we are led to the following similitude:

$$J = \Big(\frac{\tau_0}{\tau}\Big)^{4\gamma}\tilde{J}(\tilde{\tau},\sigma), \ (U,W) = U_0\Big(\frac{\tau_0}{\tau}\Big)^{\gamma}\big(\tilde{U}(\tilde{\tau},\sigma),\tilde{W}(\tilde{\tau},\sigma)\big), \ \kappa = \frac{1}{U_0\tau_0}\Big(\frac{\tau_0}{\tau}\Big)^{1-\gamma}\tilde{\kappa}(\tilde{\tau},\sigma), \tag{6}$$

where

$$\tau = -t, \quad \tilde{\tau} = -\log\frac{\tau}{\tau_0}, \quad \sigma = \frac{s_0\tau_0^{3\gamma}}{U_0\tau^{3\gamma+1}}, \tag{7}$$

and $\tau_0, U_0$ are reference scales. Because of the Beale-Kato-Majda (BKM) condition [4], we stipulate that

$$\gamma \geq 1/4. \tag{8}$$

Using (7) in (2), (3) we obtain after some manipulation

$$\tilde{J}_{\tilde{\tau}} + 4\gamma\tilde{J} + (3\gamma+1)\sigma\tilde{J}_{\sigma} + \tilde{J}(\tilde{U}\tilde{\kappa} - \bar{W}_{\sigma}) = 0, \tag{9}$$

$$(\tilde{\kappa}\tilde{J})_{\tilde{\tau}} + [(3\gamma+1)\sigma\tilde{\kappa}\tilde{J} - \bar{W}\tilde{\kappa} - \tilde{J}^{-1}\tilde{U}_{\sigma}]_{\sigma} = 0. \tag{10}$$

As we have indicated above, we shall focus in this paper on the "steady' similitude obtained by taking all variables to be independent of $\tilde{\tau}$. Unfortunately there is no evidence to our knowledge that an Euler singularity, should it occur, has this property. Indeed it has been shown that certain simpler similitudes cannot lead to such a singularity [7, 8]. However since our goal is to exhibit an example which seems to work, we apply Occam's razor here and make the simplest choice.

We shall require that the relation between $J$ and $U$ depends only upon $s_0$. Then

$$J = \frac{d\alpha}{ds_0}\left(\frac{U}{U_0}\right)^4, \tag{11}$$

where $\alpha$ is any increasing function of $s_0$. It is easily then seen that the effect is to replace $s_0$ by $\alpha$ in the similarity variable $\sigma$, which amounts to a Lagrangian relabeling. We thus choose $\alpha = s_0$.

Finally, we consider the special case $W = 0$. These assumptions yield what we shall refer to as the "basic" hairpin, which we shall use to illustrate some of the ideas involved in our approach. (We shall later restore $W$ and consider axial flow in detail.)

We now set $\tilde{U} = g(\sigma)$, $\tilde{J} = g^4$, and apply the condition $g_{\sigma}(0) = 0$ for symmetry, to obtain from (9), (10)

$$4\gamma g^4 + 4(3\gamma+1)\sigma g^3 g_{\sigma} + g^5\tilde{\kappa} = 0, \quad (3\gamma+1)\sigma\tilde{\kappa}g^4 - g^{-4}g_{\sigma} = 0. \tag{12}$$

Combining these, we have

$$\gamma(3\gamma+1)\sigma g^3 + (3\gamma+1)^2\sigma^2 g^2 g_{\sigma} + \frac{g_{\sigma}}{4g^4} = 0. \tag{13}$$

A further integration with the condition $g(0) = 1$, the latter allowed by the arbitrary constant $U_0$, yields

$$4(3\gamma+1)\sigma^2 g^{\frac{6\gamma+2}{\gamma}} + g^{\frac{2}{\gamma}} = 1. \tag{14}$$

We see that $g \sim 1 - 2\gamma(3\gamma+1)\sigma^2$ for small $\sigma$, and

$$g \sim (12\gamma+4)^{-\frac{\gamma}{6\gamma+2}}\sigma^{-\frac{\gamma}{3\gamma+1}} + O(\sigma^{-\frac{\gamma+2}{3\gamma+1}}), \tag{15}$$

for large $\sigma$. At the singularity time the velocity at the singular point is infinite, but everywhere at all previous times and away from the singular point at the singular time, the velocity is finite.

The curvature $\kappa$ of the center curve is given by

$$\kappa = -\frac{J_t}{JU}. \tag{16}$$

If $\theta$ is the angle made by $\mathbf{t}$ with the $z$-axis, so that $\kappa = \partial\theta/\partial s$, then in the similarity variables we have

$$\kappa = \frac{\tau^{\gamma-1}}{\tau_0^\gamma U_0} \frac{4\gamma g + 4(3\gamma+1)\sigma g_\sigma}{g^2}, \tag{17}$$

so that

$$\theta_\sigma = 4\gamma g^3 + 4(3\gamma+1)\sigma g^2 g_\sigma. \tag{18}$$

Using (13) this leads to

$$\theta = -\frac{1}{3\gamma+1} \int g^{-4}\sigma^{-1} dg, \tag{19}$$

and using (14) to obtain $\sigma(g)$ and integrating we find

$$\theta = \frac{2\gamma}{\sqrt{3\gamma+1}} [\pi/2 - \sin^{-1} g^{1/\gamma}]. \tag{20}$$

Thus the large-$\sigma$ asymptotes of $\mathscr{C}$ make an angle with the $z$-axis which varies from about $34°$ to $90°$ as $\gamma$ varies from $1/4$ to $1$. We thus shall refer to the center curve as the "hairpin", and the point $\sigma = 0$ on the center curve will be called the "nose" of the hairpin.

To obtain the coordinates of $\mathscr{C}$ we use $(z_s, x_s) = (\cos\theta, \sin\theta)$. Solving (14) for $\sigma g$ in terms of $g$, we can then find (from (13)) $g_\sigma$ in terms of $g$, and thus express the resulting integral for $(z, x)$ as an integral with respect to $g$. After an integration by parts we find

$$\left(\frac{\tau^{\gamma-1}}{U_0\tau_0^\gamma}\right)(z,x) = \frac{1}{\sqrt{12\gamma+4}} g \sqrt{g^{-2/\gamma}-1} [\cos\theta, \sin\theta]$$
$$+ \frac{2}{\sqrt{3\gamma+1}} \int_g^1 \sqrt{g^{-2/\gamma}-1} [\cos\theta, \sin\theta] \, dg + \frac{1}{3\gamma+1} \int_g^1 [\sin\theta, -\cos\theta] \, dg, \tag{21}$$

with $\theta(g)$ given by (20). We show the shape of the hairpin for $\gamma = 1/4$ in figure 2.

Finally, we consider the total stretching experienced by the segment of center curve defined by $(0, s_0)$, whose length is given by

$$\int_0^{s_0} J \, ds_0 = U_0 \tau_0^\gamma \tau^{1-\gamma} \int_0^\sigma g^4 \, d\sigma$$
$$\leq U_0 \tau_0^\gamma \tau^{1-\gamma} \int_0^\infty \left(g^4 - (12\gamma+4)^{-\frac{4\gamma}{6\gamma+2}} \sigma^{-\frac{4\gamma}{3\gamma+1}}\right) d\sigma + A U_0 \tau_0^\gamma \tau^{1-\gamma} \sigma^{\frac{1-\gamma}{3\gamma+1}}, \tag{22}$$

where $A = 4^{-\frac{2\gamma}{3\gamma+1}} (3\gamma+1)^{\frac{\gamma+1}{3\gamma+1}} (1-\gamma)^{-1}$, the convergence of the integral being ensured by (15). We now assume

$$1/4 \leq \gamma < 1. \tag{23}$$

In this case this last inequality bounds the total stretching as $\tau \to 0$ for fixed $s_0 > 0$ since the final term is independent of $\tau$. We thus have a situation where infinite stretching $(J = \infty)$, and simultaneously the shrinking of the dipole cross-section to zero area, occurs at a single point. The total stretching, defined as the integral of $J$ over the relevant Lagrangian segment, remains finite. Since $\theta_{g=0} = \pi/2$ when $\gamma = 1$, it is to be expected that our hairpin can actually only exist for values of $\gamma$ well below 1. In the next section we shall find it necessary to restrict $\gamma$ to $\gamma \leq 1/2$.

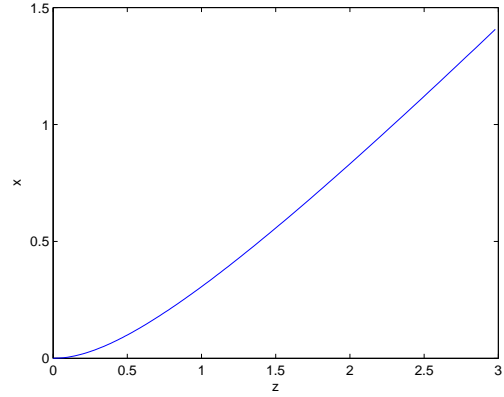

Figure 2: $\tau^{\gamma-1}/(U_0\tau_0^{\gamma})\,(z(g),x(g))$ for the basic hairpin, with $\gamma = 0.25$.

We note here for later use the equations for the center curve of the basic hairpin when we allow $\tilde{W}$ to be some known function of $\sigma$. Setting $\tilde{\kappa} = \tilde{\kappa}(\sigma)$, then (9) and the integral of (10), under the other conditions for the basic hairpin, yield

$$\tilde{\kappa} = g^{-5}[-\tilde{W}_{\sigma} + 4g^3(\gamma g + (3\gamma+1)\sigma g_{\sigma})], \tag{24}$$

$$(3\gamma+1)\sigma g^4 \tilde{\kappa} - \tilde{W}\tilde{\kappa} + g^{-4}g_{\sigma} = \text{constant} = 0, \tag{25}$$

since $\tilde{W}$ and $g_{\sigma}$ will vanish when $\sigma = 0$. Later, we shall see how to define $\tilde{W}$ from the dynamics of the axial flow of the dipole.

## 2. Fitting the snail to the center curve

### 2.1. Outline

Our aim now is to propose a construction of an Euler flow in 3D using the various components considered above. Our approach will be as follows:

- Establish a domain to contain the dipole attached to the center curve. This is a tubular domain containing a local coordinate system, within which a dipole-like vortical structure, excluding the tail, will be placed. The construction will ensure that the dipole is locally approximately two-dimensional for $\gamma \geq 1/4$.

- Carry out the construction of the snail of shrinking cross-section along the center curve. This will follow somewhat the construction in I for the axisymmetric flow without swirl, but an important non-axisymmetric effect will have to be considered, namely the axial flow resulting from variation of pressure along the center curve.

- Match with the potential flow exterior to the dipole.

We introduce the time-dependent orthogonal curvilinear coordinate system derived from the center curve, with triad $(\mathbf{n}, \mathbf{b}, \mathbf{t})$, coordinates $(\xi, \eta, \zeta)$, and metric $ds^2 = d\xi^2 + d\eta^2 + h^2 d\zeta^2$, where $h = 1 - \xi\kappa$. We adopt $\zeta$ now in place of our previous $s$ as arc length along the center curve, with $J = \partial\zeta/\partial\zeta_0$, and $d\zeta$ now refers to differential distance within a neighborhood of the center curve. The fixed coordinate system figure 1(c) applies, the $x$-coordinate pointing in the direction opposite to the dipole velocity at the nose of the hairpin.

The boundary of our tubular container will be the surface $\mathscr{C} : \xi^2 + \eta^2 = \frac{1}{4}\kappa^{-2}, -\infty < \zeta < +\infty$, the numerical factor ensuring that $h$ remains positive within the tube. We wish to solve Euler's equations within this neighborhood of the center curve, starting at an initial time $t = -\tau_0$. To do so we need to supply an initial vorticity field, then track its evolution under the constraint of Euler's equations and appropriate boundary conditions.

Recall that the center curve moves with velocity $\partial\mathbf{x}/\partial t = U\mathbf{n} + W\mathbf{t}$. Relative to our fixed frame in $\mathbb{R}^3$, the fluid velocity is $\mathbf{u} = (U + u)\mathbf{n} + v\mathbf{b} + (W + w)\mathbf{t}$. The similitude (6) applies. In our construction we shall need to introduce a characteristic diameter of the vortex dipole head, which we denote by $2a(\sigma, \tau)$. The characteristic radius $a$ is of order $U/J$ and is defined in terms of $\sigma$ by

$$a = a_0 \left( \frac{\tau}{\tau_0} \right)^{3\gamma} g^{-3}(\sigma), \quad \sigma = \frac{\zeta_0 \tau_0^{3\gamma}}{U_0 \tau^{3\gamma+1}}, \tag{26}$$

where $a_0$ is a reference radius of the dipole. Then, using (13) and (25) with the assumption $W = 0$ we have

$$a\kappa = \frac{a_0}{U_0\tau_0} \left( \frac{\tau}{\tau_0} \right)^{4\gamma-1} \widetilde{a\kappa}, \quad \widetilde{a\kappa} = \frac{4\gamma}{g^4 + 4(3\gamma+1)^2\sigma^2 g^{10}}. \tag{27}$$

We show in figure 3 the function $\widetilde{a\kappa}$ for several $\gamma$.

We are thus led to consider the important issue of quasi-two-dimensionality of the hairpin. Indeed, for fixed $\sigma$ (27) establishes local two-dimensionality if $\gamma = 1/4$ and $a_0 \ll U_0\tau_0$ and *asymptotic* local two-dimensionality as $\tau \to 0$ if $\gamma > 1/4$. We define the small ($\tau$-dependent) parameter

$$\delta = \frac{a_0}{U_0\tau_0} \left( \frac{\tau}{\tau_0} \right)^{4\gamma-1}, \quad \frac{a_0}{U_0\tau_0} \ll 1, \tag{28}$$

which will be useful for covering all cases $\gamma \geq 1/4$.

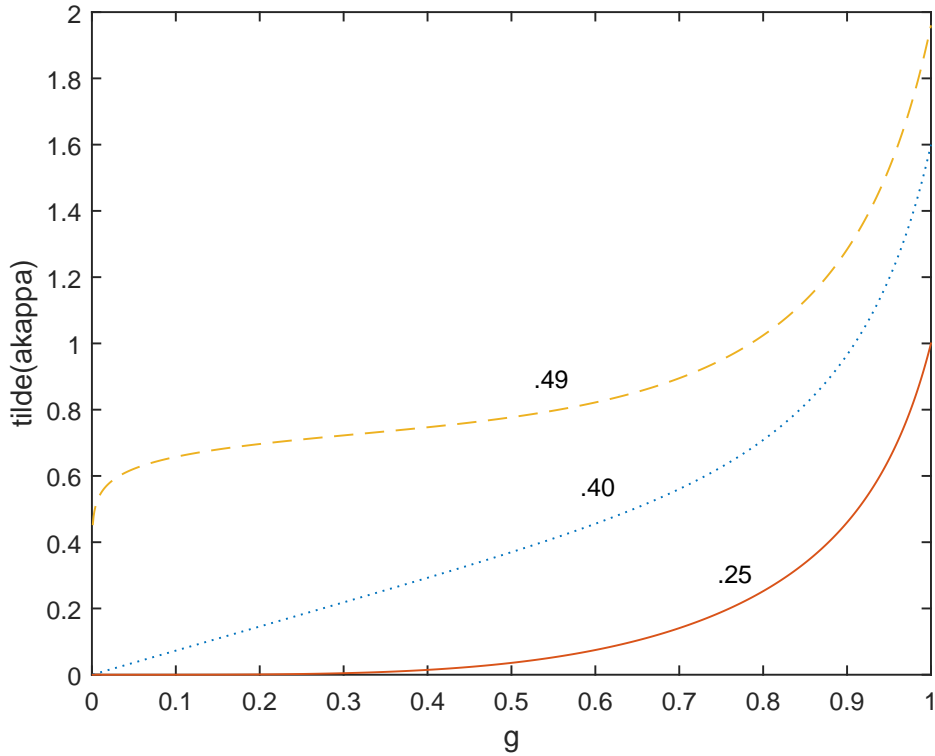

Figure 3: The function $\widetilde{a\kappa}(g)$ for, from lowest to highest, $\gamma = .25, .40, .49$.

Next consider the same issue as $\tau \to 0$ for fixed $\zeta_0$, i.e. following a Lagrangian section of the hairpin. Then, using $g \sim O(\sigma^{\frac{-\gamma}{3\gamma+1}})$ for large $\sigma$ we have

$$a\kappa \sim \left(\frac{\zeta_0}{U_0\tau_0}\right)^{4\gamma-2} \left(\frac{\tau}{\tau_0}\right) \tag{29}$$

up to a multiplicative constant. It follows that local two-dimensionality is uniform for large $\zeta_0$ provided that $1/4 \le \gamma \le 1/2$. We summarize this important feature of the hairpin as the following

**Lemma 1** *If $\frac{a_0}{U_0\tau_0} \ll 1$, then quasi two-dimensionality of the basic hairpin prevails if $\gamma = 1/4$. This feature is asymptotic for $\tau \to 0$, in the sense that $a\kappa$ vanishes with $\tau$, provided $1/4 < \gamma \le 1/2$. In both cases the property holds uniformly for large $\zeta_0$.*

So now we can describe the hairpin singularity as taking place in a domain where $\sigma$ stays $O(1)$ as $\tau \to 0$. This domain has a thickness of order $a_0(\tau/\tau_0)^{3\gamma}$ and horizontal dimensions of order $U_0\tau_0(\tau/\tau_0)^{1-\gamma}$, the ratio of the former to the latter being $\delta$. We now consider in more detail the structure within this domain.

### 2.2. Geometry of the basic snail

Since we have used AFWOS as a route to the snail, it is important to understand the structural differences between the geometry of the axisymmetric dipole and the hairpin dipole. We discuss this now for the basic snail, meaning that the snail geometry applies to the basic center curve. We will determine the characteristics of a piece of the hairpin, starting from a time $\tau = \tau_0$, and terminating at a time $\tau = 0$ at a point where vorticity at the nose is infinite but the structure at the nose has shrunk to a point. Initially, at $\tau = \tau_0$, consider a material segment of the snail at material point $\zeta_0 > 0$ on the center curve, and of axial extent $\Delta\zeta_0$. The diameter of this segment is initially

$$a_0 g^{-3}(\sigma_0), \quad \sigma_0 = \frac{\zeta_0}{U_0 \tau_0}. \tag{30}$$

The initial length of the segment is, with now $g_0 = g(\sigma_0)$,

$$L_0 = g_0^4 \Delta\zeta_0. \tag{31}$$

At later times we have,

$$a = a_0 \left(\frac{\tau}{\tau_0}\right)^{3\gamma} g^{-3}, \quad L = \left(\frac{\tau_0}{\tau}\right)^{4\gamma} g^4 L_0, \quad \sigma = \frac{\zeta_0 \tau_0^{3\gamma}}{U_0 \tau^{3\gamma+1}}. \tag{32}$$

If $A$ denotes the cross-sectional area of the snail, suitably defined with a cut-off at the tail, then we may set

$$A = A_0 \left(\frac{\tau}{\tau_0}\right)^{6\gamma} g^{-6}, \quad A_0 = a_0^2. \tag{33}$$

The volume of the segment of snail, $V$ say, is given by

$$V = V_0 \left(\frac{\tau}{\tau_0}\right)^{2\gamma} g^{-2}, \ V_0 = A_0 L_0. \tag{34}$$

The center ($\sigma = 0$) of the dipole will move (to the right say) with speed $U = -U_0(\tau_0/\tau)^\gamma$. To compute the width $H_{\text{tail}}$ of the tail as it emerges from the snail, we set

$$\frac{dV}{dt} = -\frac{dV}{d\tau} = H_{\text{tail}} U L = -H_{\text{tail}} U_0 \left(\frac{\tau_0}{\tau}\right)^\gamma g \times L_0 \left(\frac{\tau_0}{\tau}\right)^{4\gamma} g^4. \tag{35}$$

Now, taking the derivative of $V$ and using (13) we have

$$\frac{dV}{d\tau} = V_0 \frac{\tau^{2\gamma-1}}{\tau_0^{2\gamma}} \left(\frac{2\gamma}{g^2 + 4(3\gamma+1)^2 \sigma^2 g^8}\right). \tag{36}$$

Solving for $H_{\text{tail}}/a$ we obtain

$$\frac{H_{\text{tail}}}{a} = \frac{a_0}{U_0 \tau_0} \left(\frac{\tau}{\tau_0}\right)^{4\gamma-1} g^{-4} \frac{2\gamma}{1 + 4(3\gamma+1)\sigma^2 g^6}. \tag{37}$$

We thus see that we have control of the smallness of the tail thickness relative to the dipole diameter. For fixed $\sigma$ in the limit of small $\tau$ the ratio is order $\delta$ for all permissible $\gamma$. For large $\zeta_0$,

$$\frac{H_{\text{tail}}}{a} \sim \sigma_0^{\frac{4\gamma-2}{3\gamma+1}} \frac{\tau}{\tau_0} \tag{38}$$

up to a multiplicative constant, indicating that the smallness of the ratio is uniform over the hairpin for the $\gamma$ allowed in lemma 1.

Since, relative to an observer moving with a Lagrangian point on the center curve, the velocity in the tail is $O(\omega H_{\text{tail}})$, the flux of kinetic energy into the tail per unit axial length of the tail is $O(\omega^2 H_{\text{tail}}^3 U)$. We seek to compute the total volume lost per unit length of center curve,

$$\int_0^{\tau_0} \omega^2 H_{\text{tail}}^3 U \, d\tau. \tag{39}$$

Using (37) and changing the variable of integration to $\sigma$ we obtain the following estimate for the lost volume, normalized by the total initial volume per unit length, $\omega_0^2 a_0^4 g_0^{-4}$:

$$\omega_0^{-2} a_0^{-4} g_0^4 \int_0^{\tau_0} \omega^2 H_{\text{tail}}^3 U \, d\tau \sim \left(\frac{a_0}{U_0 \tau_0}\right)^2 F(\sigma_0) \int_0^{\sigma_0} G(\sigma) \, d\sigma, \tag{40}$$

where $\int_0^\infty G(\sigma) d\sigma$ exists and $F(\sigma_0)$ is bounded for $\gamma = 1/2$ and vanishes as $\sigma_0 \to \infty$ if $\gamma < 1/2$. This establishes a uniform estimate on total fractional volume lost of order $(a_0/U_0\tau_0)^2$.

The above estimates have assumed zero axial flow. How is the geometry changed by the presence of axial flow? First, some of the volume lost or gained at a given Lagrangian section can be due to axial flow. To estimate this recall that the flux through the tail per unit of $\zeta_0$ may be estimated as

$$a_0 U_0 g^2 \left(\frac{\tau_0}{\tau}\right)^{2\gamma} \frac{H_{\text{tail}}}{a} \sim \frac{a_0^2}{\tau_0} \left(\frac{\tau}{\tau_0}\right)^{2\gamma-1} g^{-2} \frac{2\gamma}{1+4(3\gamma+1)\sigma^2 g^6}. \tag{41}$$

On the other hand the flux due to variation of axial flow with $\zeta_0$ can be estimated as follows. We have already taken $W = (\tau_0/\tau)^\gamma \tilde{W} = (\tau_0/\tau)^\gamma h(\sigma)$ and we shall see in section 4 that this ordering applies to the axial velocity in general. Thus we have the estimate of this variation as

$$\frac{\partial}{\partial \zeta_0} \left[ U_0 a_0^2 \left(\frac{\tau}{\tau_0}\right)^{5\gamma} h g^{-6} \right] \sim \frac{a_0^2}{\tau_0} \left(\frac{\tau}{\tau_0}\right)^{2\gamma-1} (h g^{-6})_\sigma. \tag{42}$$

If both $h$ and $g$ are $O(\sigma^{\frac{-\gamma}{3\gamma+1}})$ for large $\sigma$ then these two estimates may be shown to be uniform at large $\zeta_0$ provided that $\gamma \leq 1/2$.

Regarding the assumed asymptotic behaviour of $g$ and $h$ we point out that the error term in (15) changes in the presence of axial flow, to $O(\sigma^{-\frac{\gamma+1}{3\gamma+1}})$. Then, for large $\sigma$ the two centerline equations are approximated by

$$h_\sigma \approx 4g^3(\gamma g + (3\gamma+1)\sigma g_\sigma), \quad (3\gamma+1)\sigma k + g^{-8} g_\sigma \approx 0, \tag{43}$$

which is consistent with $h = O(\sigma^{\frac{-\gamma}{3\gamma+1}})$.

Another feature of the axial flow is associated with the flux of vorticity into the wake. Because of the axial variation of circulation in each eddy of the dipole, some vortex lines are turned downstream and trail into the wake as $x$-directed vorticity. To compute this flux we note that $a^2 \omega \sim g^{-2}(\tau/\tau_0)^2 a_0^2 \omega_0$. The $\zeta_0$ derivative is equal to the flux through the piece of the tail defined by $\Delta\zeta_0$, namely $H_{\text{tail}}(\tau_0/\tau)^4 g^4 w_y$. Integrating the equality from $y = 0$ to $y = H_{\text{tail}}/2$ we get the dimensionless jump in axial flow across half of the shed tail as

$$U_0^{-1}\Delta w = -g^{-7}g_\sigma(\tau_0/\tau)^\gamma\delta. \tag{44}$$

The factor $\delta$ means this effect is small compared to the dipole velocity.

### 2.3. Dynamics of the hairpin

In Cartesian coordinates Euler's equations for an incompressible fluid are

$$\mathbf{u}_t + \mathbf{u}\cdot\nabla\mathbf{u} + \nabla p = 0, \ \ \nabla\cdot\mathbf{u} = 0. \tag{45}$$

We need to compute acceleration of a fluid particle relative to our moving frame attached to the moving center curve. Now, if $\mathbf{x}(\mathbf{x}_0,t)$ is the position of a fluid particle $\mathbf{x}_0$, we have

$$\frac{\partial}{\partial t}\Big|_{\mathbf{x}_0} = \frac{\partial}{\partial t}\Big|_{\mathbf{x}} + (U+u)\frac{\partial}{\partial\xi} + v\frac{\partial}{\partial\eta} + h^{-1}(W+w)\frac{\partial}{\partial\zeta}$$

$$\equiv \frac{D}{Dt} + u\frac{\partial}{\partial\xi} + v\frac{\partial}{\partial\eta} + h^{-1}w\frac{\partial}{\partial\zeta} \equiv \mathscr{D}. \tag{46}$$

Also

$$\frac{\partial(\mathbf{n},\mathbf{b},\mathbf{t})}{\partial t}\Big|_{\mathbf{x}_0} = (-h^{-1}U_\zeta\mathbf{t}, 0, h^{-1}U_\zeta\mathbf{n}), \ \frac{\partial(\mathbf{n},\mathbf{b},\mathbf{t})}{\partial\xi} = 0, \ \frac{\partial(\mathbf{n},\mathbf{b},\mathbf{t})}{\partial\eta} = 0, \tag{47}$$

and

$$\frac{\partial(\mathbf{n},\mathbf{b},\mathbf{t})}{\partial\zeta} = (-\kappa\mathbf{t}, 0, \kappa\mathbf{n}). \tag{48}$$

Using these relations in (45), including now the pressure term $\nabla p = (p_\xi, p_\eta, h^{-1}p_\zeta)$, and gathering components we obtain

$$\mathscr{D}u + 2h^{-1}(W+w)\frac{\partial U}{\partial\zeta} + h^{-1}(W+w)^2\kappa + \frac{\partial p}{\partial\xi} + \frac{\partial U}{\partial t} = 0, \tag{49}$$

$$\mathscr{D}v + \frac{\partial p}{\partial\eta} = 0, \tag{50}$$

$$\mathscr{D}w + h^{-1}w\frac{\partial W}{\partial\zeta} - h^{-1}u\frac{\partial U}{\partial\zeta} - h^{-1}(W+w)(U+u)\kappa$$

$$+ h^{-1}\frac{\partial(p - \frac{1}{2}U^2)}{\partial\zeta} + \frac{\partial W}{\partial t} + h^{-1}W\frac{\partial W}{\partial\zeta} = 0. \tag{51}$$

The equation $\nabla\cdot\mathbf{u} = 0$ becomes

$$\frac{\partial h(U+u)}{\partial\xi} + \frac{\partial hv}{\partial\eta} + \frac{\partial(W+w)}{\partial\zeta} = 0. \tag{52}$$

Recalling that $h_\xi = -\kappa$, this may be written

$$\frac{\partial u}{\partial\xi} + \frac{\partial v}{\partial\eta} + h^{-1}\frac{\partial(W+w)}{\partial\zeta} = h^{-1}\kappa(U+u). \tag{53}$$

The vorticity vector is

$$(\omega_n, \omega_b, \omega_t) \equiv (\omega_n, \omega_b, \omega) = \Big(\frac{\partial w}{\partial \eta} - \frac{1}{h}\frac{\partial v}{\partial \zeta}, h^{-1}\frac{\partial (U+u)}{\partial \zeta} - \frac{\partial w}{\partial \xi} + h^{-1}\kappa(W+w), \frac{\partial v}{\partial \xi} - \frac{\partial u}{\partial \eta}\Big).$$
$$(54)$$

From (49) and (50) we calculate the following equation for the axial vorticity $\omega_t \equiv \omega$:

$$\mathscr{D}\omega + h^{-1}(U+u)\kappa\omega - h^{-1}\frac{\partial (W+w)}{\partial \zeta}\omega + h^{-2}\kappa(W+w)\frac{\partial v}{\partial \zeta}$$

$$+h^{-1}\frac{\partial w}{\partial \xi}\frac{\partial v}{\partial \zeta} - h^{-1}\frac{\partial w}{\partial \eta}\frac{\partial u}{\partial \zeta} - 2h^{-1}\frac{\partial w}{\partial \eta}\frac{\partial U}{\partial \zeta} - 2h^{-1}\kappa(W+w)\frac{\partial w}{\partial \eta} = 0. \qquad (55)$$

We remark that we have introduced the uniform component $W$ of the axial flow without indicating how this splitting is to be made. The point is that the uniform component appears in the "uniform" material derivative $D/Dt$. We shall later find it necessary to define $W$ to be the axial flow along the center of the eddies of the dipole, but for the moment this fact is unimportant.

We rearrange (55), (51), and (53) as follows (recall $\omega$ is now the tangential component of vorticity):

$$u\omega_\xi + v\omega_\eta = F_\omega, \qquad (56)$$

$$uw_\xi + vw_\eta = F_w, \qquad (57)$$

$$u_\xi + v_\eta = F. \qquad (58)$$

From the geometry of the hairpin described in the preceding section, we can show that the "forcing terms" on the RHS of (56), (57) and (58), suitably normalized, are $O(\delta)$ relative to the LHS, where again $\delta = a_0/(U_0\tau_0)(\tau/\tau_0)^{4\gamma-1}$. We shall consider this in further detail below, where we make use of the contour average

$$\langle \cdot \rangle = \oint \frac{\cdot}{\sqrt{u^2 + v^2}}\, ds \qquad (59)$$

taken over a closed streamline of the unperturbed dipole eddies. (Given the up-down symmetry we focus always on the upper eddy.) The compatibility conditions obtained under this contour averaging have the form $\langle F_\omega \rangle = \langle F_w \rangle = 0$. Note that the streamfunction in the cross-flow plane is here taken as defined by $(u, v) = (-\psi_\eta, \psi_\xi)$.

### 2.4. The basic singularity with $w = W = 0$

We introduce now a simplified model using the basic center curve and snail, which neglects all axial flow, $w = W = 0$. With this assumption we have in (56), (58),

$$F_\omega = -\omega_t - h^{-1}(U+u)\kappa\omega, \quad F = h^{-1}(U+u)\kappa. \qquad (60)$$

Here the time derivative is the material derivative for a particle moving normal to the center curve with velocity $U$. In this model the local vorticity structure will be quasi-2D and will respond only to the stretching induced by the normal motion of the dipole in the presence of curvature. Following the discussion of the axisymmetric case, we shall impose local conservation of energy to determine the choice of lateral scale, with $a$ given by (26). This leads to the ordering and expansion to be discussed next. We shall see that in the absence of axial flow we can avoid contour averaging to obtain leading order compatibility.

With our ordering the vorticity equation (56) has the form, using the fact that $h = 1 + O(\delta)$,

$$u\omega_\xi + v\omega_\eta = -\omega_t - [1 + O(\delta)](U + u)\kappa\omega. \tag{61}$$

The LHS of this is $O(\omega U/a)$ while the RHS is $O(\omega\tau)$. The RHS is smaller by a factor $a_0/U_0\tau_0$, or more appropriately (once similarity variables are introduced) by a factor $\delta$, than the LHS. Thus we have

$$u\omega_\xi + v\omega_\eta = -\omega_t - (U + u)\kappa\omega, \tag{62}$$

with an error $O(\delta^2)$ relative to the left-hand side. Setting $\omega = (U_0/a_0)J\tilde{\omega}$ ($\tilde{\omega}$ then being dimensionless), and using (2), we obtain the approximate vorticity equation

$$u\tilde{\omega}_\xi + v\tilde{\omega}_\eta = -\tilde{\omega}_t - u\kappa\tilde{\omega}. \tag{63}$$

Also we have the approximate continuity equation from (58), (60),

$$u_\xi + v_\eta = (U + u)\kappa. \tag{64}$$

We now need to appropriately renormalize the remaining variables. Setting $(u, v) = |U|(\tilde{u}, \tilde{v})$ and $(\xi, \eta) = a(\tilde{\xi}, \tilde{\eta})$, we see that

$$\frac{\partial \tilde{\omega}}{\partial t}\Big|_{\xi,\eta} = \frac{\partial \tilde{\omega}}{\partial t}\Big|_{\tilde{\xi},\tilde{\eta}} + \frac{a_\tau}{a}\left[\tilde{\xi}\frac{\partial \tilde{\omega}}{\partial \tilde{\xi}} + \tilde{\eta}\frac{\partial \tilde{\omega}}{\partial \tilde{\eta}}\right]. \tag{65}$$

Recalling the time variable defined by $\bar{\tau} = -\log\tau$ we have

$$\frac{\partial}{\partial t} = -\frac{\partial}{\partial \tau} = \frac{1}{\tau}\frac{\partial}{\partial \bar{\tau}}. \tag{66}$$

Thus, using the fact that $a/(|U|\tau) = \delta g^{-4}$, we set

$$(\tilde{u}, \tilde{v}) = (u^\dagger, v^\dagger) + \delta g^{-4}\frac{a_{\bar{\tau}}}{a}(\tilde{\xi}, \tilde{\eta}), \quad \kappa = \delta a^{-1}\tilde{\kappa}, \tag{67}$$

and the vorticity equation becomes

$$u^\dagger\frac{\partial \tilde{\omega}}{\partial \tilde{\xi}} + v^\dagger\frac{\partial \tilde{\omega}}{\partial \tilde{\eta}} = -\delta\left[g^{-4}\frac{\partial \tilde{\omega}}{\partial \bar{\tau}} + \tilde{u}\tilde{\kappa}\tilde{\omega}\right]. \tag{68}$$

The continuity equation becomes, neglecting a term of order $\delta^2$,

$$\frac{\partial u^\dagger}{\partial \tilde{\xi}} + \frac{\partial v^\dagger}{\partial \tilde{\eta}} = \delta g^{-4}\frac{J_{\bar{\tau}}}{J}(\tfrac{1}{2} + u^\dagger) = \delta g^{-5}[4\gamma g + 4(3\gamma + 1)\sigma g_\sigma](\tfrac{1}{2} + u^\dagger). \tag{69}$$

We now take the leading order term $\tilde{\omega}_0$ to be independent of $\bar{\tau}$, so that

$$u^\dagger\frac{\partial \tilde{\omega}}{\partial \tilde{\xi}} + v^\dagger\frac{\partial \tilde{\omega}}{\partial \tilde{\eta}} = -\delta\tilde{u}\tilde{\kappa}\tilde{\omega}. \tag{70}$$

Thus

$$\left[ u^\dagger \frac{\partial}{\partial \tilde{\xi}} + v^\dagger \frac{\partial}{\partial \tilde{\eta}} \right] \left[ \tilde{\omega}(1 + \delta \tilde{\xi} \tilde{\kappa}) \right] = O(\delta^2). \tag{71}$$

Although there are slight differences between the problem posed by (70) and (71) and that treated in section 3 of I, it is seen that the result is the same provided the contour integration $\langle \cdot \rangle$ uses $(u^\dagger, v^\dagger)$ neglecting terms of order $\delta$. That is, we can take $u^\dagger, v^\dagger$ including terms of order $\delta$, and from (70) conclude that $\tilde{\omega}$ is to leading order constant on the spiral streamlines, hence constant everywhere, or else we may use contour integration, with $(u^\dagger, v^\dagger)$ divergence free to leading order.

From either viewpoint we see, writing $\tilde{\omega} = \tilde{\omega}_0 + O(\delta)$, that *the only compatible solution (to order $\delta$ inclusive) is $\tilde{\omega}_0 = constant$.* Since the propagation velocity of the vortex is now unity, and if $\tilde{A} \equiv A/a^2$ is the dimensionless upper eddy area, we know that we are dealing, to leading order, with a Sadovskii vortex, and therefore must have $\tilde{\omega}_0^2 \tilde{A}_0 = 37.11$ from [31]; see also [30].

### 2.5. *The potential flow exterior to the hairpin*

In the axisymmetric problem studied in I, we gave a detailed account of the exterior potential flow and the shedding of vorticity into the tail which is associated with it. That analysis verified the *ad hoc* assumptions of the snail model. In the problem at hand, we are dealing with a complex three-dimensional structure. In principle there is an analogous potential flow problem by which we could test the assumptions of our model. We will not, however, attempt this calculation here, but we will consider some of the features of this exterior flow.

In the first place, we can estimate pressure forces which might need to be considered as boundary conditions on the hairpin flow. One component is the free stream flow past the Sadovskii vortex. The associated pressure is (returning to primitive variables) of order $U^2$. Since the vortex is also accelerating in the normal direction, i.e. $U_t \neq 0$, there is an inertial force associated with its apparent mass. The associated pressure is $O(aU_t)$. But $aU_t/U^2 \sim a/(U\tau) = O(\delta)$ and so this is negligible. Another component of the local exterior flow is a sink-like flow caused by the contraction of vortex area beyond that due to conservation of volume. Near the vortex it is a flow of order $\delta$, with associated pressures of order $\delta^2$, and so it is again negligible.

Next, it is necessary to verify the local two-dimensionality of the potentials. The potentials we need to consider are, first, the potential dipole distribution associated with local flow over the vortical dipole, second, the distribution of sinks associated with the shrinkage of the dipole beyond that due to conservation of volume, and third, the potential flow which matches with the axial flow at the snail boundary. All three lead to similar estimates, and the analysis below will deal only with the second. For the first we can appeal to the results of I for potential flow over an expanding torus, which indicates that the effect of non-locality on the dipole is of order $\delta \log \delta$ and so is negligible.

We will show that a sink distribution of the form

$$\tilde{\phi}(\mathbf{x}) \equiv \frac{\phi}{|U|a} = \frac{1}{4\pi} \int_{\mathscr{C}} f(\zeta_0^{\mathbf{y}}, \tau) J(\zeta_0^{\mathbf{y}}) |\mathbf{x} - \mathbf{y}|^{-1} d\zeta_0^{\mathbf{y}} \tag{72}$$

acts locally due to the approximate two-dimensionality of the hairpin dipole. Going over to local coordinates near a point $\zeta_0^{\mathbf{x}}$ on the center line, we consider a point $\mathbf{x}$ on the circle $r^2 = \xi^2 + \eta^2$ in the plane orthogonal to the center curve at $\zeta_0^{\mathbf{x}}$, with $\mathbf{y}$ the Lagrangian point $\zeta_0^{\mathbf{y}}$ on the center curve. We need to show that when $r/a \gg 1$ and $r\kappa \ll 1$ we have at $\mathbf{x}$ a local evaluation of $\xi\tilde{\phi}_\xi + \eta\tilde{\phi}_\eta$ at the point $\zeta_0$. Since $a\kappa = O(\delta)$ we may for example take $a/r = O(\delta^{1/2}), r\kappa = O(\delta^{1/2})$.

We first consider the integral restricted to the segment $(\zeta^{\mathbf{x}} - \varepsilon, \zeta^{\mathbf{x}} + \varepsilon)$ or $(\zeta_0^{\mathbf{x}} - J^{-1}(\zeta_0^{\mathbf{x}})\varepsilon, \zeta_0^{\mathbf{x}} + J^{-1}(\zeta_0^{\mathbf{x}})\varepsilon)$ where $\varepsilon/r \sim \delta^{-1/2}$. Then, in (72), we have

$$|\mathbf{x} - \mathbf{y}|^{-1} \approx \left(B^2(\sigma^{\mathbf{x}})(\sigma^{\mathbf{x}} - \sigma^{\mathbf{y}})^2 + r^2\right)^{-1/2}, \quad B(\sigma) = U_0\tau_0(\tau/\tau_0)^{1-\gamma}g^4. \tag{73}$$

The integral in question can then be brought into the form

$$-\frac{f(\sigma^{\mathbf{x}})}{4\pi}\int_{-\varepsilon/r}^{+\varepsilon/r}\frac{dx}{(x^2+1)^{3/2}} \approx -\frac{f(\sigma^{\mathbf{x}})}{4\pi}\int_{-\infty}^{+\infty}\frac{dx}{(x^2+1)^{3/2}} = -\frac{f(\sigma^{\mathbf{x}})}{4\pi}. \tag{74}$$

By matching with the dipole we then have

$$\xi\tilde{\phi}_\xi + \eta\tilde{\phi}_\eta = -\frac{f(\sigma^{\mathbf{x}})}{4\pi} + \cdots = \frac{A}{4\pi|U|a\tau}\frac{J_{\tilde\tau}}{J} + \cdots, \tag{75}$$

where the dots signify the error from the remainder of the integral of the center curve. We consider only the piece from $\zeta^{\mathbf{x}}$ to $\infty$, denoted by $\mathscr{I}$. The integral from the nose and the integral over the branch not containing $\mathbf{x}$ can be estimated similarly.

We thus have to estimate, up to multiplicative constants,

$$|\mathscr{I}| \sim \int_{\sigma^{\mathbf{x}}+\sigma_\varepsilon}^{\infty} f(\sigma^{\mathbf{y}})\frac{Br^2\,d\sigma^{\mathbf{y}}}{\left[(\int_{\sigma^{\mathbf{x}}}^{\sigma^{\mathbf{y}}}B(\sigma)d\sigma)^2 + r^2\right]^{3/2}}, \quad \sigma_\varepsilon = J^{-1}(\sigma^{\mathbf{x}})\left(\frac{\tau_0}{\tau}\right)^{3\gamma+1}\frac{\varepsilon}{U_0\tau_0}. \tag{76}$$

Clearly

$$|\mathscr{I}| < \int_{\sigma^{\mathbf{x}}+\sigma_\varepsilon}^{\infty} f(\sigma^{\mathbf{y}})\frac{Br^2\,d\sigma^{\mathbf{y}}}{\left[\int_{\sigma^{\mathbf{x}}}^{\sigma^{\mathbf{y}}}B(\sigma)d\sigma\right]^3}. \tag{77}$$

From the behavior of $g$ for large $\sigma$ the integral is easily shown to converge, but it diverges at the lower limit of integration, as

$$\frac{f(\sigma^{\mathbf{x}})r^2}{B^2\sigma_{\tilde\varepsilon}^2} = \frac{r^2}{\varepsilon^2} \sim \delta. \tag{78}$$

Thus we find that the distant contributions of the sink distribution are negligible and the shrinkage of the snail acts locally on the exterior flow.

Finally, there is the axial flow on the bounding streamline of the dipole. This would need to be matched with a locally 2D potential flow, in the form of a Dirichlet problem for its potential determined by the constant axial derivative of the potential around the dipole boundary. The effect of the tail has been disregarded in all of these potential flow problems.

### 3. The full dynamics with axial flow

#### 3.1. Compatibility

We now give the right-hand sides of the equations (56), (57), (58) which will enter into our analysis. We omit terms which will not contribute to the compatibility constraints obtained below, that is, which will vanish under expansion in $\delta$ and contour averaging. (We note in particular that $\langle u \rangle = \langle v \rangle = 0$ when the flow $(u, v)$ is divergence free and has closed streamlines.) These are

$$F'_\omega = -\frac{D\omega}{Dt} - U\kappa\omega - w\omega_\zeta + (w + W)_\zeta \omega - w_\xi v_\zeta + w_\eta u_\zeta, \tag{79}$$

$$F'_w = -\frac{D(W + w)}{Dt} - w(w + W)_\zeta + (w + W)U\kappa - (p - \tfrac{1}{2}U^2)_\zeta, \tag{80}$$

$$F' = \kappa U - (w + W)_\zeta. \tag{81}$$

We now give the result of applying the compatibility procedure of [9]. To leading order with the forcing terms absent, we see that we have a divergence-free 2D flow with $w$ and $\omega$ constant on the closed streamlines of the unperturbed dipole. The compatibility conditions result from the contour average $\langle \cdot \rangle$ on these streamlines applied to (56) and (57) with use of (58). To understand how (58) is employed, consider (56) and expand as $\omega = \omega_0 + \omega_1 + \ldots$ where the first-order term $\omega_1$ is of the order of $F_\omega^0$, the RHS evaluated on the zeroth-order terms. The contour averaging leads to $\langle \mathbf{u}_1 \cdot \nabla \omega_0 \rangle = \langle \mathbf{u}_1 \cdot \mathbf{n}\, \partial \omega_0 / \partial \psi \rangle = \langle F_\omega^0 \rangle$. The divergence theorem then brings in (58). For details see [9].

We thus obtain the following two compatibility equations:

$$A_\psi[D_\psi\omega - (W_\zeta - U\kappa)\omega] - \omega \frac{\partial(w, A)}{\partial(\zeta, \psi)} + w_\psi \int \frac{\partial(H_\psi, A)}{\partial(\zeta, \psi)}\, d\psi = 0, \tag{82}$$

$$A_\psi[D_\psi(w + W) - U\kappa(w + W)] + \frac{\partial(H, A)}{\partial(\zeta, \psi)} - \int \frac{\partial(H_\psi, A)}{\partial(\zeta, \psi)}\, d\psi - UU_\zeta A_\psi = 0, \tag{83}$$

where

$$D_\psi \equiv \frac{\partial}{\partial t}\Big|_{\psi, \zeta_0} + w\frac{\partial}{\partial \zeta} + \mathscr{V}\frac{\partial}{\partial \psi}, \tag{84}$$

and

$$\mathscr{V} = -\frac{1}{A_\psi}\Big[A_t + wA_\zeta + \int \frac{\partial(w, A)}{\partial(\zeta, \psi)}\, d\psi - (U\kappa - W_\zeta)A\Big]. \tag{85}$$

We note that (83) may also be written

$$A_\psi[D_\psi(w + W) - U\kappa(w + W) + (H - U^2/2)_\zeta] = \Gamma_\zeta, \tag{86}$$

where $\Gamma = \int H_\psi A_\psi\, d\psi$, taken from the eddy center to a contour $\psi = \psi^*$, is the circulation at contour $\psi^*$.

We can also write the compatibility equations in the following alternate forms:

$$A_\psi D_\psi\omega + A_\psi U\kappa\omega + \frac{\partial(w + W, \Gamma)}{\partial(\psi, \zeta)} = 0, \tag{87}$$

$$A_\psi D_\psi(w + W) - A_\psi U \kappa(w + W) + \omega^{-1} \frac{\partial(\Gamma, H - \frac{1}{2}U^2)}{\partial(\psi, \zeta)} = 0. \tag{88}$$

From these last two equations we then have the relation for the helicity density,

$$A_\psi D_\psi[\omega(w + W)] + \frac{\partial(\Gamma, H - \frac{1}{2}[U^2 + (w + W)^2])}{\partial(\psi, \zeta)} = 0. \tag{89}$$

The unknowns here may be taken as $\omega$, $w$, $A$, $\mathcal{V}$ and $H$, all taken as functions of $\zeta, \psi, \tau$ and governed by the equations (82), (83), (85), plus $H_\psi = \omega$ and $U, W, \kappa$ as functions of $\zeta, \tau$ and satisfying the two center curve equations (2), (3) (with $J = (U/U_0)^4$). The missing two equations are, first, $\nabla^2 \psi = \omega$, which must be solved to obtain $A$ as a function of $\psi$, and second, a relation determining $U$ from the dipole structure (cf. $\omega^2 A/U^2 = 37.11$ for the Sadovskii snail). This system is thus of the form of a so-called generalized differential equation, where the coordinate streamlines must be determined as a side condition; see [20]. A procedure similar to that applied in our analysis of the case of zero axial flow may now be used to study the snail in the shrinking coordinates.

### 3.2. The axial driving force

We focus now on the mechanism of generation of axial flow for the contour averaged equations. From (86) we may identify the averaged axial pressure gradient as a force $-\langle \mathcal{F} \rangle$ applied to the fluid, where

$$\langle \mathcal{F} \rangle = (H - U^2/2)_\zeta - A_\psi^{-1} \Gamma_\zeta. \tag{90}$$

We consider now some properties of $\mathcal{F}$.

**Lemma 2** *For the zeroth-order Sadovskii vortex, i.e the dipole of leading order as $\tau \to 0$, having constant $\omega$ over each eddy, the forcing function $\mathcal{F}$ given by (90) has average zero, in the sense that $\int_{\text{eddy}} \langle \mathcal{F} \rangle A_\psi \, d\psi = 0$, if and only if the eddy has the scaling of the Sadovskii snail.*

To prove this, let $\psi_c$ be the value of $\psi$ at the center of the eddy, $\psi$ being zero at the outer boundary. For a Sadovskii eddy, $H - U^2/2 = \omega \psi$ where $\omega$ is independent of $\psi$. Using the fact that the $\zeta$-derivatives are taken holding $\psi$ fixed, we have

$$\int_{\psi_c}^0 \langle \mathcal{F} \rangle A_\psi \, d\psi = -\frac{\partial}{\partial \zeta} \int_{\psi_c}^0 \Gamma \, d\psi + \int_{\psi_c}^0 A_\psi (\omega \psi)_\zeta \, d\psi = \frac{\partial}{\partial \zeta} \int_0^{\Gamma_0} \psi \, d\Gamma + \frac{\omega_\zeta}{\omega} \int_0^{\Gamma_0} \psi \, d\Gamma, \tag{91}$$

where we have used that $\Gamma = 0$ when $\psi = \psi_c$ and $\psi = 0$, $\Gamma = \Gamma_0$ at the outer boundary of the eddy. This expression will equal zero if and only if $\int_0^{\Gamma_0} \psi \, d\Gamma$ is a multiple of $\omega^{-1}$. Now, in terms of scalings in $J$, $\psi \Gamma \sim U^2 a^2 \sim \omega^2 a^4$. If this is to scale like $\omega^{-1}$, we must have $a \sim \omega^{-3/4} \sim J^{-3/4}$, which is the scaling of the snail.

We can check this lemma for the case of a circular eddy. There $\omega \psi = \omega^2(r^2 - a^2)/2$ and $A_\psi = 2\pi/\omega$. Thus $\int_{\text{eddy}} \langle \mathcal{F} \rangle A_\psi \, d\psi$ is proportional to

$$\frac{2\pi}{g^4} \int_0^{g^3} \left[ \frac{2\pi}{g^4}(4g^7 g_\zeta r^2 - gg_\zeta) - 4\pi g^3 g_\zeta r^2 \right] r \, dr = 0. \tag{92}$$

Another result of interest is

**Lemma 3** *If the dipole has the similarity of the hairpin and $\langle \mathscr{F} \rangle$ is independent of $\psi$, then necessarily, as functions of $\psi$ for any $\zeta$, $\Gamma$ is a constant multiple $C$ of $A_\psi^{1/2}$ and $\langle \mathscr{F} \rangle = C(d/d\zeta)[A_\psi(\zeta, \psi_{\text{center}})]^{-1/2}$, $C$ being positive in the upper eddy.*

To show this let $H^* = H - \frac{1}{2}U^2$. If $H_\zeta^* - A_\psi^{-1}\Gamma_\zeta$ has zero derivative with respect to $\psi$ then, since the $\zeta$ derivatives may be taken holding $\psi$ fixed, a $\psi$ derivative gives

$$\omega_\zeta + A_\psi^{-2} A_{\psi\psi}\Gamma_\zeta - A_\psi^{-1}(A_\psi\omega)_\zeta = A_\psi^{-2}\frac{\partial(A_\psi,\Gamma)}{\partial(\psi,\zeta)} = 0. \tag{93}$$

Now the similarity of the hairpin requires

$$\psi = g^2\tilde{\psi}, \quad A_\psi(\zeta,\psi) = g^{-4}\tilde{A}_{\tilde{\psi}}(g^2\tilde{\psi}), \quad \Gamma(\zeta,\psi) = g^{-2}\tilde{\Gamma}(g^2\tilde{\psi}). \tag{94}$$

It follows that

$$\Gamma = CA_\psi^{1/2}. \tag{95}$$

It turns out that $A_\psi$ tends to be almost constant in many propagating dipoles. Since a constant circulation is not an option, we must deal with an $\langle \mathscr{F} \rangle$ which varies from streamline to streamline.

### 3.3. Erosion

In order to understand the approach we now need to implement, it is helpful to recall the methods used in the axisymmetric case without swirl. In I we basically obtained (11) heuristically by *assuming* that erosion was occurring, in a way consistent with local conservation of energy. The basic assumption was then that the eddy dimension $a$ was of order $R^{-3/4}$ where $R$ was the radius of the center curve of the dipole. In the general case $J$ replaces $R$ and so $a \sim (\tau/\tau_0)^{3/4}$ must be our starting point. We also showed in I that a full asymptotic treatment of the erosion process led from first principles to the ordering $R^{-3/4}$.

Thus we commence our analysis with $a \sim (\tau/\tau_0)^{3/4}$, not (11). We deal with the primitive equations where terms which vanish under contour integration are neglected, along with higher-order terms in the small parameter $\delta$ of (28):

$$u\omega_\xi + v\omega_\eta = \frac{\partial\omega}{\partial\tau} - w\frac{\partial\omega}{\partial\zeta} - U\kappa\omega + \omega\frac{\partial(w+W)}{\partial\zeta} - \frac{\partial w}{\partial\xi}\frac{\partial v}{\partial\zeta} + \frac{\partial w}{\partial\eta}\frac{\partial u}{\partial\zeta}, \tag{96}$$

$$uw_\xi + vw_\eta = \frac{\partial(w+W)}{\partial\tau} - w\frac{\partial(w+W)}{\partial\zeta} + U\kappa(w+W) - \frac{\partial}{\partial\zeta}(p - \frac{1}{2}U^2), \tag{97}$$

$$u_\xi + v_\eta = U\kappa - (w+W)_\zeta. \tag{98}$$

Rather than introduce $g(\sigma)$ to enforce the strict similarity of the basic hairpin, we define a general set of similarity variables. A complete list follows:

$$(\tilde{\xi}, \tilde{\eta}) = (\tau/\tau_0)^{-3\gamma}a_0^{-1}(\xi,\eta), \quad \tilde{\omega}(\sigma,\tilde{\xi},\tilde{\eta},\tilde{\tau}) = \omega_0^{-1}(\tau/\tau_0)^{4\gamma}\omega, \tag{99}$$

$$(\tilde{u}, \tilde{v}, \tilde{w})(\sigma, \tilde{\xi}, \tilde{\eta}, \tilde{\tau}) = (\tau/\tau_0)^\gamma U_0^{-1}(u, v, w), \ \ \tilde{J}(\sigma, \tilde{\tau}) = (\tau/\tau_0)^{4\gamma} J, \tag{100}$$

$$\tilde{\psi}(\sigma, \tilde{\xi}, \tilde{\eta}, \tilde{\tau}) = (\tau/\tau_0)^{-2\gamma} U_0^{-1} a_0^{-1} \psi, \ \ (\tilde{U}, \tilde{W})(\sigma, \tilde{\tau}) = U_0^{-1}(\tau/\tau_0)^\gamma(U, W), \tag{101}$$

$$\tilde{p}(\sigma, \tilde{\xi}, \tilde{\eta}, \tilde{\tau}) = U_0^{-2}(\tau/\tau_0)^{2\gamma} p, \ \ \sigma = U_0 \tau_0^{-1} \zeta_0 (\tau/\tau_0)^{-(3\gamma+1)}, \ \ \tilde{\tau} = -\log(\tau/\tau_0), \tag{102}$$

$$\tilde{\kappa}(\sigma_{\tilde{\tau}}) = U_0 \tau_0 (\tau/\tau_0)^{1-\gamma} \kappa, \ \ \tilde{\Gamma}(\sigma, \tilde{\psi}, \tilde{\tau}) = \omega_0^{-1} a_0^{-2} (\tau_0/\tau)^{2\gamma} \Gamma, \ \ \tilde{A}(\sigma, \tilde{\psi}, \tilde{\tau}) = a_0^{-1}(\tau_0/\tau)^{6\gamma} A. \tag{103}$$

We are free to set $a_0 \omega_0 = U_0$. Introducing these variables into our primitive equations we have

$$(\tilde{u} + 3\gamma\delta\tilde{\xi})\tilde{\omega}_{\tilde{\xi}} + (\tilde{v} + 3\gamma\delta\tilde{\eta})\tilde{\omega}_{\tilde{\eta}} = \delta[-\tilde{\omega}_{\tilde{\tau}} - (3\gamma+1)\sigma\tilde{\omega}_\sigma - \tilde{w}\tilde{J}^{-1}\tilde{\omega}_\sigma - \tilde{\omega}\tilde{J}^{-1}(\tilde{w} + \tilde{W})_\sigma$$
$$+ \tilde{\omega}[\tilde{J}^{-1}\tilde{J}_{\tilde{\tau}} + (3\gamma+1)\tilde{J}^{-1}\sigma\tilde{J}_\sigma] - \tilde{J}^{-1}\tilde{w}_{\tilde{\xi}}\tilde{v}_\sigma + \tilde{J}^{-1}\tilde{w}_{\tilde{\eta}}\tilde{u}_\sigma], \tag{104}$$

$$(\tilde{u} + 3\gamma\delta\tilde{\xi})\tilde{w}_{\tilde{\xi}} + (\tilde{v} + 3\gamma\delta\tilde{\eta})\tilde{w}_{\tilde{\eta}} = \delta[-\gamma(\tilde{w} + \tilde{W}) - (\tilde{w} + \tilde{W})_{\tilde{\tau}} - (3\gamma+1)\sigma(\tilde{w} + \tilde{W})_\sigma$$
$$- \tilde{J}^{-1}\tilde{w}\tilde{w}_\sigma + \tilde{J}^{-1}\tilde{W}\tilde{W}_\sigma + (\tilde{w} + \tilde{W})[\tilde{J}^{-1}\tilde{J}_{\tilde{\tau}} + 4\gamma + (3\gamma+1)\tilde{J}^{-1}\sigma\tilde{J}_\sigma]$$
$$- \tilde{J}^{-1}(\tilde{H} - \tilde{U}^2/2)_\sigma + \tilde{J}^{-1}(\tilde{u}\tilde{u}_\sigma + \tilde{v}\tilde{v}_\sigma)], \tag{105}$$

$$\tilde{u}_{\tilde{\xi}} + \tilde{v}_{\tilde{\eta}} = \delta[-4\gamma - \tilde{J}^{-1}(3\sigma+1)\sigma\tilde{J}_\sigma - \tilde{J}^{-1}\tilde{J}_{\tilde{\tau}} - \tilde{J}^{-1}\tilde{w}_\sigma]. \tag{106}$$

Here all $\tilde{\tau}$ derivatives are taken for fixed $\sigma, \tilde{\xi}, \tilde{\eta}$. Other relations which follow are

$$\tilde{\psi}_{\tilde{\xi}} = \tilde{v}, \ \ \tilde{\psi}_{\tilde{\eta}} = -\tilde{u}, \ \ \tilde{\Gamma} = \int \tilde{\omega}\tilde{A}_{\tilde{\psi}} \, d\tilde{\psi}, \ \ \int \tilde{w}\tilde{A}_{\tilde{\psi}} \, d\tilde{\psi} = 0, \ \ \tilde{\psi}_{\tilde{\xi}\tilde{\xi}} + \tilde{\psi}_{\tilde{\eta}\tilde{\eta}} = \tilde{\omega}. \tag{107}$$

Now (2) and (3) yield

$$\tilde{J}_{\tilde{\tau}} + 4\gamma\tilde{J} + (3\gamma+1)\sigma\tilde{J}_\sigma + \tilde{J}(\tilde{U}\tilde{\kappa} - \tilde{W}_\sigma) = 0, \tag{108}$$

$$(\tilde{\kappa}\tilde{J})_{\tilde{\tau}} + [(3\gamma+1)\sigma\tilde{\kappa}\tilde{J} - \tilde{W}\tilde{\kappa} - \tilde{J}^{-1}\tilde{U}_\sigma]_\sigma = 0. \tag{109}$$

Performing contour integration to find the compatibility equations in tilde variables, we obtain:

$$\tilde{A}_{\tilde{\psi}}\Big[\tilde{J}[(\tilde{\omega}/\tilde{J})_{\tilde{\tau}} + (3\gamma+1)\sigma(\tilde{\omega}/\tilde{J})_\sigma] + \tilde{J}^{-1}\tilde{w}\tilde{\omega}_\sigma + \tilde{\mathcal{V}}\tilde{\omega}_{\tilde{\psi}}\Big]$$
$$- \tilde{J}^{-1}\tilde{\omega}\frac{\partial(\tilde{w}, \tilde{A})}{\partial(\sigma, \tilde{\psi})} + \tilde{J}^{-1}\tilde{w}_{\tilde{\psi}}\int \frac{\partial(\tilde{\omega}, \tilde{A})}{\partial(\sigma, \tilde{\psi})} \, d\tilde{\psi} = 0, \tag{110}$$

$$\tilde{A}_{\tilde{\psi}}[(\tilde{w} + \tilde{W})_{\tilde{\tau}} + \gamma(\tilde{w} + \tilde{W}) + (3\gamma+1)\sigma(\tilde{w} + \tilde{W})_\sigma + \tilde{w}(\tilde{w} + \tilde{W})_\sigma + \tilde{\mathcal{V}}\tilde{w}_{\tilde{\psi}}$$
$$- \tilde{U}\tilde{\kappa}(\tilde{w} + \tilde{W}) + \tilde{J}^{-1}(\tilde{H} - \tfrac{1}{2}\tilde{U}^2)_\sigma] - \tilde{J}^{-1}\tilde{\Gamma}_\sigma = 0. \tag{111}$$

Here

$$\tilde{\mathcal{V}} = -\tilde{A}_{\tilde{\psi}}^{-1}\Big[\tilde{A}_{\tilde{\tau}} + (3\gamma+1)\sigma\tilde{A}_\sigma + \tilde{w}\tilde{J}^{-1}\tilde{A}_\sigma \tag{112}$$
$$+ \tilde{J}^{-1}\int \frac{\partial(\tilde{w}, \tilde{A})}{\partial(\sigma, \tilde{\psi})} \, d\tilde{\psi} + A[\tilde{J}^{-1}\tilde{J}_{\tilde{\tau}} - 2\gamma + \tilde{J}^{-1}(3\gamma+1)\sigma\tilde{J}_\sigma]\Big].$$

### 3.4. The $\tilde{\tau}$-independent problem

Dropping $\tilde{\tau}$ derivatives and removing the tildes we have the following two dynamical equations:

$$A_\psi \left[ J(3\gamma+1)\sigma(\omega/J)_\sigma + J^{-1}w\omega_\sigma + \mathscr{V}\omega_\psi \right] - J^{-1}\omega\frac{\partial(w,A)}{\partial(\sigma,\psi)} + J^{-1}w_\psi \int \frac{\partial(\omega,A)}{\partial(\sigma,\psi)}\,d\psi = 0, \tag{113}$$

$$A_\psi[\gamma(w+W) + (3\gamma+1)\sigma(w+W)_\sigma + w(w+W)_\sigma + \mathscr{V}w_\psi$$
$$- U\kappa(w+W) + J^{-1}(H - \tfrac{1}{2}U^2)_\sigma] - J^{-1}\Gamma_\sigma = 0. \tag{114}$$

Here

$$\mathscr{V} = -A_\psi^{-1}\left[(3\gamma+1)\sigma A_\sigma + wJ^{-1}A_\sigma + J^{-1}\int\frac{\partial(w,A)}{\partial(\sigma,\psi)}\,d\psi + A[-2\gamma + J^{-1}(3\gamma+1)\sigma J_\sigma]\right]. \tag{115}$$

### 3.5. Symmetrization

We shall now replace the variable $\psi$ by the variable $A$. To do this we make use repeatedly of the chain rule

$$\frac{\partial(\cdot)}{\partial\sigma}\Big|_\psi = \frac{\partial(\cdot)}{\partial\sigma}\Big|_A + \frac{\partial(\cdot)}{\partial A}\Big|_\sigma A_\sigma, \tag{116}$$

and also

$$A_\psi^{-1}\frac{\partial(\cdot)}{\partial\psi}\Big|_\sigma = \frac{\partial(\cdot)}{\partial A}\Big|_\sigma. \tag{117}$$

Using these we find that some of the terms involving $\mathscr{V}\partial_\psi$ combine with $\sigma$ derivatives at fixed $\psi$ to yield $\sigma$ derivatives at fixed $A$ through (116). We then obtain from (113)

$$J^2\mathscr{L}(\Gamma_A/J) + \frac{\partial(Q,\Gamma_A)}{\partial(A,\sigma)} + \frac{\partial(Q_A,\Gamma)}{\partial(A,\sigma)} = 0, \tag{118}$$

where

$$\mathscr{L} = (3\gamma+1)\sigma\frac{\partial}{\partial\sigma} + \left[2\gamma - J^{-1}(3\gamma+1)\sigma J_\sigma\right]A\frac{\partial}{\partial A}, \tag{119}$$

and

$$\Gamma = \int\omega\,dA, \quad Q = \int w\,dA. \tag{120}$$

Similarly, for (114) we find

$$\gamma(Q_A+W) + (3\gamma+1)\sigma(Q_{A\sigma}+W_\sigma) + Q_A(Q_{A\sigma}+W_\sigma) - Q_{AA}\left[J^{-1}Q_\sigma + A(-2\gamma + J^{-1}(3\gamma+1)\sigma J_\sigma)\right]$$
$$- U\kappa(Q_A+W) + J^{-1}\left[(H-\tfrac{1}{2}U^2)_\sigma - A_\psi^{-1}\Gamma_\sigma\right] = 0. \tag{121}$$

Here of course all $\sigma$ derivatives are at fixed $A$. The appearance of $H$ and $A_\psi$ complicates this expression. Otherwise, given the functions of $\sigma$ from the center curve, and $W$, we would have two equations in $\Gamma, Q$. But both $H$ and $A_\psi$ come from solving our dipole problem $\nabla^2\psi = \Gamma_A = \omega$. Note that $A_\psi H_A = \Gamma_A$ is another relation that can be used.

It is interesting that (118) allows an integration with respect to $A$, yielding

$$J^2 \mathscr{L}(\Gamma/J) + \frac{\partial(Q,\Gamma)}{\partial(A,\sigma)} - J[2\gamma - J^{-1}(3\gamma+1)\sigma J_\sigma]\Gamma = 0. \tag{122}$$

The system is completed with

$$4\gamma J + (3\gamma+1)\sigma J_\sigma + J(U\kappa - W_\sigma) = 0, \tag{123}$$

$$(3\gamma+1)\sigma\kappa J - W\kappa - J^{-1}U_\sigma = 0. \tag{124}$$

We can check that we regain our basic hairpin for the case of zero axial flow. If $Q = W = 0$ (122) is solved by $\Gamma = \text{constant} \times AJ$. We then impose conservation of energy by $\omega^2 A^2 J \sim \omega^3 A^2 \sim 1$ so $A \sim \omega^{-3/2}$ and $\omega A^{1/2} \sim \omega^{1/4}$. Setting $J = g^4(\sigma)$, $U = g(\sigma)$, $A \sim a^2 = g^{-6}$ creates the previous singular hairpin.

### 3.6. Helicity

The local helicity density is defined as $\omega \cdot \mathbf{u}$. Consider the helicity balance within a fixed volume $V$ bounded by a closed surface $\partial V$. Then, by the equations of motion,

$$\frac{d}{dt}\int_V \omega \cdot \mathbf{u}\, dV = \int_V (\omega_t \cdot \mathbf{u} + \omega \cdot \mathbf{u}_t)\, dt$$
$$= -\int_{\partial V}[\mathbf{u}\,\omega \cdot \mathbf{u} + \omega(p - \tfrac{1}{2}(u^2 + v^2 + w^2)^2)] \cdot \mathbf{n}\, dS. \tag{125}$$

The helicity balance follows that of volume, with the additional factor of order $U^2\omega \sim (\tau_o/\tau)^{5\gamma}$. Thus

$$\frac{d}{dt}\int_V \omega \cdot \mathbf{u}\, dV = O(\tau_0/\tau)^{3\gamma+1}, \tag{126}$$

and this is balanced by axial and tail fluxes as computed in section 2.2, the volume fluxes there being proportional to $(\tau/\tau_0)^{2\gamma-1}$.

## 4. Preliminary analysis of the system

We begin by recapitulating our problem. We have the general equations

$$J^2 \mathscr{L}(\Gamma/J) + \frac{\partial(Q,\Gamma)}{\partial(A,\sigma)} - J[2\gamma - J^{-1}(3\gamma+1)\sigma J_\sigma]\Gamma = 0, \tag{127}$$

where

$$\mathscr{L} = (3\gamma+1)\sigma\frac{\partial}{\partial\sigma} + \left[2\gamma - J^{-1}(3\gamma+1)\sigma J_\sigma\right]A\frac{\partial}{\partial A}, \tag{128}$$

$$\gamma(Q_A + W) + (3\gamma+1)\sigma(Q_{A\sigma} + W_\sigma) + Q_A(Q_{A\sigma} + W_\sigma) - Q_{AA}\left[J^{-1}Q_\sigma + A(-2\gamma + J^{-1}(3\gamma+1)\sigma J_\sigma)\right]$$
$$- U\kappa(Q_A + W) + J^{-1}\left[(H - \tfrac{1}{2}U^2)_\sigma - A_\psi^{-1}\Gamma_\sigma\right] = 0. \tag{129}$$

For the center line equations we have

$$4\gamma J + (3\gamma+1)\sigma J_\sigma + J(U\kappa - W_\sigma) = 0, \tag{130}$$

$$(3\gamma+1)\sigma\kappa J - W\kappa - J^{-1}U_\sigma = 0. \tag{131}$$

Let us rewrite (127) and (129) as follows:

$$\mathscr{L}_1(\Gamma/J) = -J^{-2}\frac{\partial(Q,\Gamma)}{\partial(A,\sigma)} \equiv \mathscr{F}_1, \tag{132}$$

$$\mathscr{L}_2(Q_A+W) + J^{-1}\big[(H-\tfrac{1}{2}U^2)_\sigma - A_\psi^{-1}\Gamma_\sigma\big] = J^{-1}[WW_\sigma - Q_A Q_{A\sigma} + Q_{AA}Q_\sigma] \equiv \mathscr{F}_2, \tag{133}$$

where

$$\mathscr{L}_1 = \mathscr{L} - (2\gamma - J^{-1}(3\gamma+1)\sigma J_\sigma), \quad \mathscr{L}_2 = \mathscr{L} + (5\gamma + J^{-1}(3\gamma+1)\sigma J_\sigma). \tag{134}$$

*4.1. Effect of axial flow on the center curve*

We now outline how we shall estimate the effect of axial flow on the hairpin geometry. First of all we may set $J = g(\sigma)^4$ and $U = -g(\sigma) + \Delta U$. Then (130) and (131) may be combined to yield the following equation:

$$g^{2-2/\gamma}\Big[2\gamma(3\gamma+1)\sigma^2 g^{\frac{6\gamma+2}{\gamma}} + \frac{\gamma}{2}g^{\frac{2}{\gamma}}\Big]_\sigma + \mathscr{E} = 0, \tag{135}$$

where

$$\mathscr{E} = \frac{(-g\Delta U + \tfrac{1}{2}\Delta U^2)_\sigma (3\gamma+1)\sigma g^4 + Wg g_\sigma}{(3\gamma+1)\sigma g^4 - W} - g^8 W_\sigma(3\gamma+1)\sigma. \tag{136}$$

Thus we may determine the distorted $\sigma(g)$ from

$$\sigma^2 = \frac{1 - g^{2/\gamma} + \frac{2}{\gamma}\int_g^1 g^{2/\gamma-2}\mathscr{E}g_\sigma^{-1}dg}{4(3\gamma+1)g^{\frac{6\gamma+2}{\gamma}}}. \tag{137}$$

This distortion can therefore be determined if $\Delta U$ and $W$ are known. In general we do not know $\Delta U$ until the distribution of vorticity is known and the local two-dimensional propagating dipole with this distribution, if it exists, has been determined. We shall see that $W$ can be calculated approximately and that it is necessarily the axial flow at the center of the dipole eddies. All of these quantities involve $\sigma$, so that $\sigma(g)$ is here understood to exist. We shall see below that (137) can be put into the form of a differential equation which can be solved numerically.

*4.2. Approximations*

We propose to carry out an approximate calculation of axial flow and vorticity distribution, in several steps. We first neglect $\mathscr{F}_1$ and $\mathscr{F}_2$. We then solve (138) for $\Gamma$. Since axial flow has been temporarily expelled, this must yield the Sadovskii eddy. Next, we solve (139) for $W$ and $Q_A$, with $\mathscr{F}_2$ zero, using the known $\Gamma$. Then we compute a next approximation to $\Gamma$ by restoring $\mathscr{F}_1$ and using the just calculated values of $W$ and $Q_A$. If $\Delta U$ can be determined from the improved $\Gamma$, then it, together with $W$, may be used in (137) to compute the shape of the hairpin. We then check the importance of neglected terms in the construction.

It will turn out that some remarkable properties of the approximate solution enable the shape of the basic hairpin to be only slightly distorted in the presence of axial flow, provided that $\Delta U$ and the vorticity distribution are suitably chosen. The question then is whether or not such a dipole can be realized as a 2D Euler flow. We shall return to this question below.

### 4.3. A first approximation to the symmetrized equations

We shall approximate our problem by initially setting $\mathscr{F}_{1,2} = 0$. We thus have the following equations:

$$\mathscr{L}_1(\Gamma/J) = 0, \tag{138}$$

$$\mathscr{L}_2(Q_A + W) + J^{-1}\big[(H - \tfrac{1}{2}U^2)_\sigma - A_\psi^{-1}\Gamma_\sigma\big] = 0. \tag{139}$$

Consider first (138). Now

$$\mathscr{L}_1 = (3\gamma+1)\sigma\frac{\partial}{\partial\sigma} + \big[2\gamma - J^{-1}(3\gamma+1)\sigma J_\sigma\big]A\frac{\partial}{\partial A} - (2\gamma - J^{-1}(3\gamma+1)\sigma J_\sigma). \tag{140}$$

In this operator we neglect axial flow effects on $J$ and thus we use the function appropriate to the basic hairpin, namely $J = g^4$. Then the solution of (138) is seen to be

$$\Gamma/J = \sigma^{\frac{2\gamma}{3\gamma+1}}g^{-4}F(Ag^4\sigma^{\frac{-2\gamma}{3\gamma+1}}), \tag{141}$$

where, if an unphysical singularity is to be avoided at $\sigma = 0$ as well as at $A = 0$, we must have $F(x) = C_0 x$. As expected, we recover the Sadovskii structure of constant vorticity eddies. For the Sadovskii dipole we take the constant $C_0$ as $\sqrt{37.11}$ and the eddy area as $g^{-6}$ to obtain $U = -g$ as the first approximation to $U$. Thus

$$\Gamma = \sqrt{37.11}\,AJ = \sqrt{37.11}\,Ag^4. \tag{142}$$

Next, consider the axial flow. We see that it is being forced by $J^{-1}\big[(H - \tfrac{1}{2}U^2)_\sigma - A_\psi^{-1}\Gamma_\sigma\big]$. We now compute this forcing term for the Sadovskii dipole. The numerical calculation of the dipole with these parameters (see section 5) then yields, at the nose of the hairpin,

$$A = 3.634\psi + 1. \tag{143}$$

Then

$$H - \tfrac{1}{2}U^2 = \omega\psi = \sqrt{37.11}(0.275)(A-1) = 1.675(A-1) \equiv C(A-1). \tag{144}$$

Away from the nose these become

$$H - \tfrac{1}{2}U^2 = Cg^2(Ag^6 - 1), \quad g^6A = 3.634g^2\psi + 1. \tag{145}$$

Also

$$\Gamma = \omega A = \sqrt{37.11}A, \tag{146}$$

which becomes, away from the nose,

$$\Gamma = \sqrt{37.11}g^4 A. \tag{147}$$

Thus

$$\left[H - \tfrac{1}{2}U^2\right]_\sigma - \Gamma_\sigma/A_\psi = Cgg_\sigma(4g^6 A - 2). \tag{148}$$

Note that the integral of the RHS with respect to $A$ from 0 to $g^{-6}$ gives zero, consistent with lemma 1 when applied to the symmetrized problem.

From equation (139) for $Q_A + W$ we have

$$\mathscr{L}_2(Q_A + W) = -J^{-1}\left[(H - \tfrac{1}{2}U^2)_\sigma - A_\psi^{-1}\Gamma_\sigma\right] = Cg^{-3}g_\sigma(2 - 4g^6 A). \tag{149}$$

This equation is straightforward to solve, and the unique solution, regular at $\sigma = 0$, is

$$Q_A + W = -a(\sigma) + b(\sigma)A/A_0, \tag{150}$$

where

$$a(\sigma) = C\,\frac{2\sigma^{\frac{-5\gamma}{3\gamma+1}}}{(3\gamma+1)g^4} \int_g^1 g\sigma^{\frac{2\gamma-1}{3\gamma+1}}\,dg, \tag{151}$$

$$b(\sigma) = C\,\frac{4\sigma^{\frac{-7\gamma}{3\gamma+1}}}{(3\gamma+1)g^6} \int_g^1 g^3\sigma^{\frac{4\gamma-1}{3\gamma+1}}\,dg. \tag{152}$$

Here $a(\sigma)$ is a local function not related to the lateral length scale appearing earlier in the text. Note that to determine these functions of $\sigma$ we need to make use of a connection between $g$ and $\sigma$, which at this stage of the iteration, we take to be that of the basic center curve:

$$\sigma = \frac{1}{2\sqrt{3\gamma+1}}\,g^{-\frac{3\gamma+1}{\gamma}}\sqrt{1 - g^{\frac{2}{\gamma}}}. \tag{153}$$

Next, we restore $\mathscr{F}_1$ and solve

$$\mathscr{L}_1(\Gamma/J) = J^{-2}(Q_\sigma\Gamma_A - Q_A\Gamma_\sigma), \tag{154}$$

where we need to use (146) and an expression for $Q$. Now we know that $Q_A = -(a + W) + b(\sigma)Ag^6$ and hence (since $Q$ vanishes at the eddy center) $Q = -(a + W)A + (b/2)g^6 A^2$. Substitutions on the right hand side of (154) then lead to a forcing term proportional to $A[(a + W)/J]_\sigma$. We can show that $a$ is proportional to $\sigma$ near $\sigma = 0$, and we require for regularity that $W = O(\sigma)$ there. Thus this forcing term will in general lead to a singularity like $\log\sigma$, unless $W = -a$. We are forced to define $W$ to be the axial flow along the center of the dipole eddy. That our Lagrangian variable is linked to the center of the eroding dipole seems natural as the structure collapses.

With now $W = -a$, $Q = (b/2)g^6 A^2$, (154) becomes

$$\mathscr{L}_1(\Gamma/J) = \sqrt{37.11}\,d(\sigma)A^2, \quad d = J\left(\frac{bg^6}{2J^2}\right)_\sigma = g^4\left(\frac{b}{2g^2}\right)_\sigma. \tag{155}$$

It is again straightforward to solve this equation. The first correction to $\Gamma = \sqrt{37.11}AJ$ is thus given by

$$\Gamma = \sqrt{37.11}J[A + D(\sigma)A^2], \tag{156}$$

where

$$D = \frac{8C}{3\gamma+1}(1-g^{\frac{2}{\gamma}})^{-\frac{\gamma}{3\gamma+1}}g^6[I_1 + I_2 - I_3], \tag{157}$$

$$I_1 = \int_g^1 (1-x^{\frac{2}{\gamma}})^{-\frac{2\gamma+1}{3\gamma+1}}x^{\frac{2-\gamma}{\gamma}}\,dx, \tag{158}$$

$$I_2 = \int_g^1 (1-x^{\frac{2}{\gamma}})^{\frac{-8\gamma+1}{6\gamma+2}}\int_x^1 y^{\frac{1-\gamma}{\gamma}}(1-y^{\frac{2}{\gamma}})^{\frac{4\gamma-1}{6\gamma+2}}\,dy\,dx, \tag{159}$$

$$I_3 = \frac{7}{3\gamma+1}\int_g^1 x^{\frac{3-\gamma}{\gamma}}(1-x^{\frac{2}{\gamma}})^{-\frac{14\gamma+3}{6\gamma+2}}\int_x^1 y^{\frac{1-\gamma}{\gamma}}(1-y^{\frac{2}{\gamma}})^{\frac{4\gamma-1}{6\gamma+2}}\,dy\,dx. \tag{160}$$

### 4.4. Approximate conservation of energy

We now indicate a surprising property of this calculation. Recall that we are engaged in essentially an iterative approximation where we compute axial flow generated within the Sadovskii dipole, and then use this flow to determine the modified vorticity distribution. Since we are not solving the system exactly, there is no guarantee that the scalings associated with local conservation of energy are maintained. If they are, then $D$ should be proportional to $g^6$, and this is certainly not obvious from the expression for $D$ above. However it turns out that in fact we have to a good approximation

$$D = 4C\frac{3\gamma+1}{10\gamma+1}g^6. \tag{161}$$

Here the expression on the right comes from evaluation of the exact expression as $g \to 1$. In figure 4 we compare the exact $D$ with this approximation for $\gamma = .25, .5$. The good agreement allows us to use this simplification in subsequent calculations.

### 4.5. Calculation of the distorted hairpin

We now return to (136) and (137) to determine $g(\sigma)$ in the presence of axial flow with the above approximations. We first note that the derivative of $\sigma(g)$ at $g = 1$ can be extracted exactly from (136), (137). Setting $\sigma^2 \approx r(1-g)$ near $g = 1$ and $\Delta U = -\delta Ug$, we find that $r$ satisfies

$$r = \frac{1}{2\gamma(3\gamma+1)}\Big[1 + \frac{(3\gamma+1)(2\delta u + \delta u^2)r - \frac{4C}{8\gamma+1}}{(3\gamma+1)r + \frac{4C}{8\gamma+1}} - \frac{6\gamma+2}{8\gamma+1}C\Big]. \tag{162}$$

The existence of a real $r$ will be an important constraint on solutions.

We define $X = g$ and the 2-vector $Y(X) = (\sigma^2 g^{\frac{6\gamma+1}{\gamma}}, \int_g^1 g\sigma^{\frac{2\gamma-1}{3\gamma+1}}dg)$. Then (135) and (136) (with $W = -a(\sigma)$), yield a system of two equations of the form,

$$\frac{dY}{dX} = \mathscr{F}(X, Y) \tag{163}$$

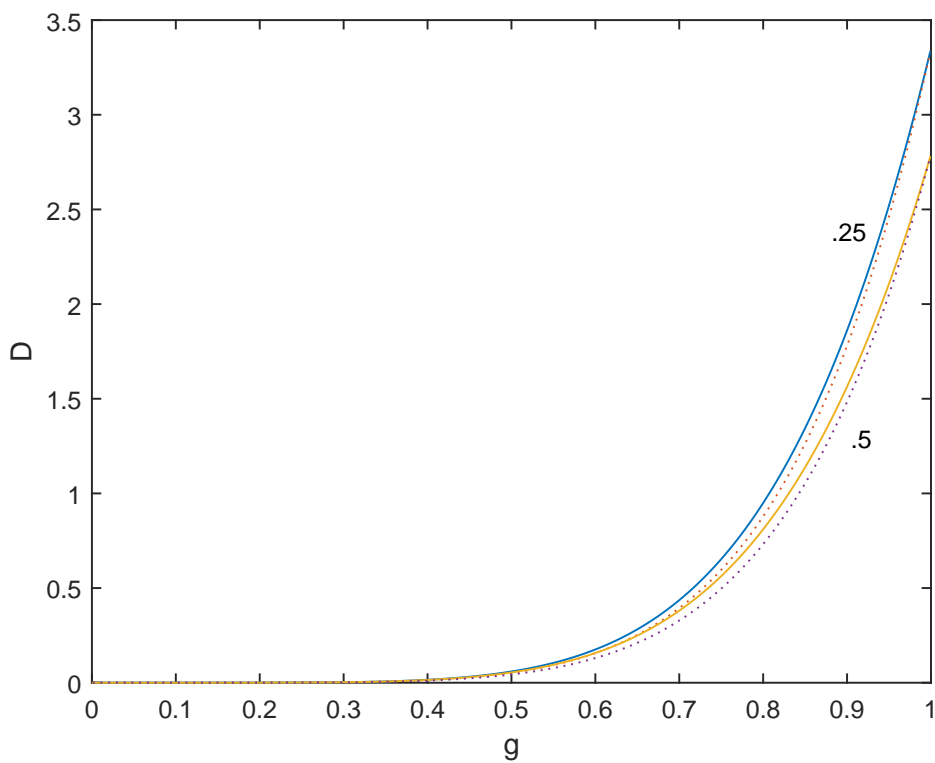

Figure 4: Exact functions $D(g)$ (solid) compared with (161) (dotted), for $\gamma = .25, .5$.

We may solve this system numerically given slightly displaced initial values near $X = 1$ and the relevant value of $r$. For sufficiently large $\delta U$ two positive values of $r$ exist, but only the larger value yields acceptable behaviour at $\sigma = \infty$.

For $\gamma = 1/4, \delta U = 1.5$ we show $Y_1(g)$ in figure 5, compared with two approximations proportional to $1 - g^8$. Both of these approximations correspond to a simple relabeling of the basic hairpin, obtained by either matching the asymptotic form for large $\sigma$, or else fitting to the derivative at $g = 1$. Real values of $r$ cease to exist for slightly smaller values of $\delta U$. If we adopt the matching at $\sigma = \infty$ we obtain a $g$ close to a relabeling of the basic center curve.

### 4.6. Dipole structure

The crucial issue is now whether or not these approximate calculations are leading toward parameter values achievable by a locally 2D propagating dipole. Because of the structure of our iteration from the Sadovskii case it is convenient to standardize the description of these dipoles. As we shall see in the next section, there is a family of extensions of the Sadovskii structure to dipoles with vorticity varying from streamline to streamline but constant on each streamline. To a good approximation, both vorticity and the area $A$ are linear functions of $\psi$. Since this is also true of the perturbed Sadovskii dipole which we have just calculated, we may take as our family

$$\omega = \sqrt{37.11}(1 + c^* A / A_0^*). \tag{164}$$

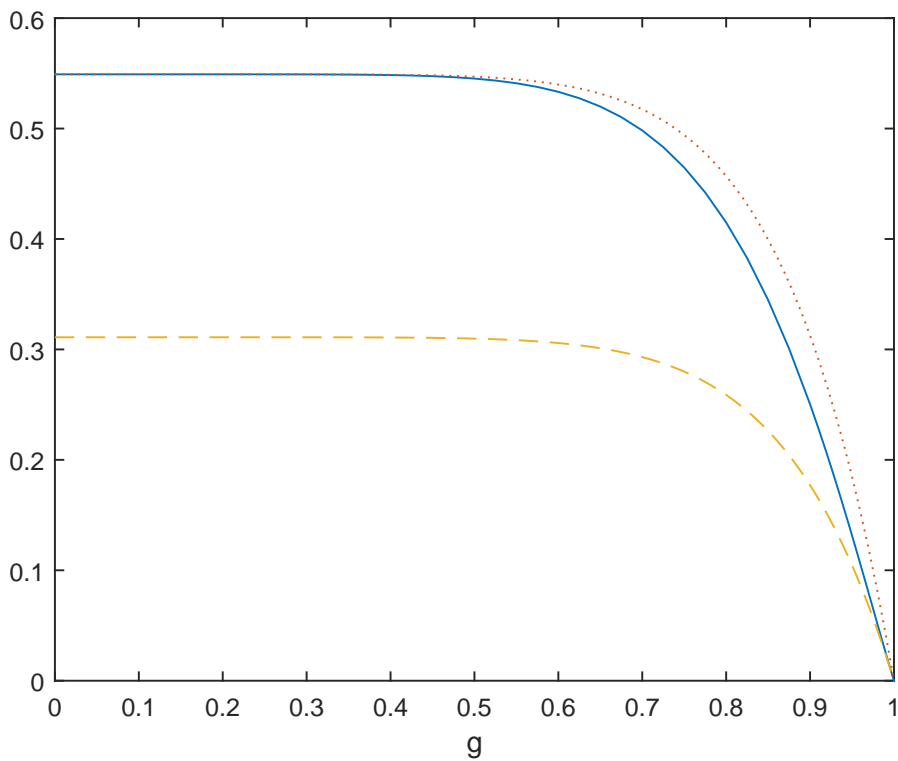

Figure 5: $Y_1(g)$ for $\gamma = .25$, $\delta U = 1.5$ obtained by solving (163) (solid), $.549(1 - g^8)$ (dotted), $.311(1 - g^8)$ (dashed).

Here $A_0^*$ is the area of one eddy. We associate with any such dipole a translation velocity $U^*$.

Let us take the case of figure 5 as an example. When we used the $\sigma(g)$ of the basic hairpin to compute our constant $D$ (recall $\omega = \sqrt{37.11}(1 + 2DA)$, we obtained $D = 2C$ for $\gamma = 1/4$. However the $Y_1$ in figure 5 is well approximated by

$$\sigma^2 = .549 g^{14}(1 - g^8) = K \times \tfrac{1}{7} g^{14}(1 - g^8), \quad K = 3.843. \tag{165}$$

It is not hard to see that if $\sigma^2$ is altered by a factor $K$, then $D$ is changed to $D/K$. Since $4C/K = 4 \times 1.675/3.843 = 1.74$, we see that for this calculation $c^* = 1.74$. Also $A_0^* = 1$ and $U^* = 2.5$. In the following table we indicate analogous calculations for various $c^*$ for two values of $\gamma$. Note that for values of $-U^*$ slightly smaller than 2.5, there is no real value of $r$, and so such a steady propagating dipole is not compatible with the hairpin, according to these approximate calculations.

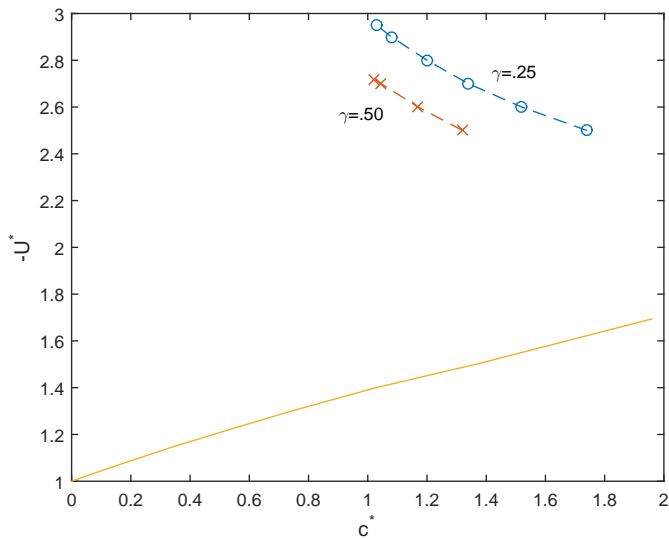

Figure 6: Required values of $c^*, -U^*$ for two values of $\gamma$, according to the approximate computation of axial flow, compared with the calculations of section 5 for a family of propagating dipoles (solid line).

| $\gamma = .25$ | | $\gamma = .5$ | |
|---|---|---|---|
| $c^*$ | $-U^*$ | $c^*$ | $-U^*$ |
| 1.74 | 2.5 | 1.32 | 2.5 |
| 1.51 | 2.6 | 1.17 | 2.6 |
| 1.34 | 2.7 | 1.045 | 2.7 |
| 1.20 | 2.8 | 1.0226 | 2.72 |
| 1.08 | 2.9 | | |
| 1.03 | 2.95 | | |

In the next section we compute a family of propagating dipoles with vorticity a linear function of the streamfunction. We thus have here a family of *obtainable* values of $c^*, U^*$ in this class of dipoles. In figure 6 we compare these computed values with those of the above table. It is clear that we cannot at present match the required dipole with our available dipoles. Whether or not this becomes possible with a better approximation (there is some indication that $-U^*$ is lowered in the next iteration for axial flow) or through an enlarged family of 2D solutions obtained by a different method, remains to be seen. There is also of course the possibility that the fault rests with the assumption of steady similitude. At this juncture it appears that the most likely candidate for preventing a finite time singularity of steady type is the breakup of the dipole as a result of axial flow. It is also conceivable that a succession of instabilities might occur as the dipole erodes.

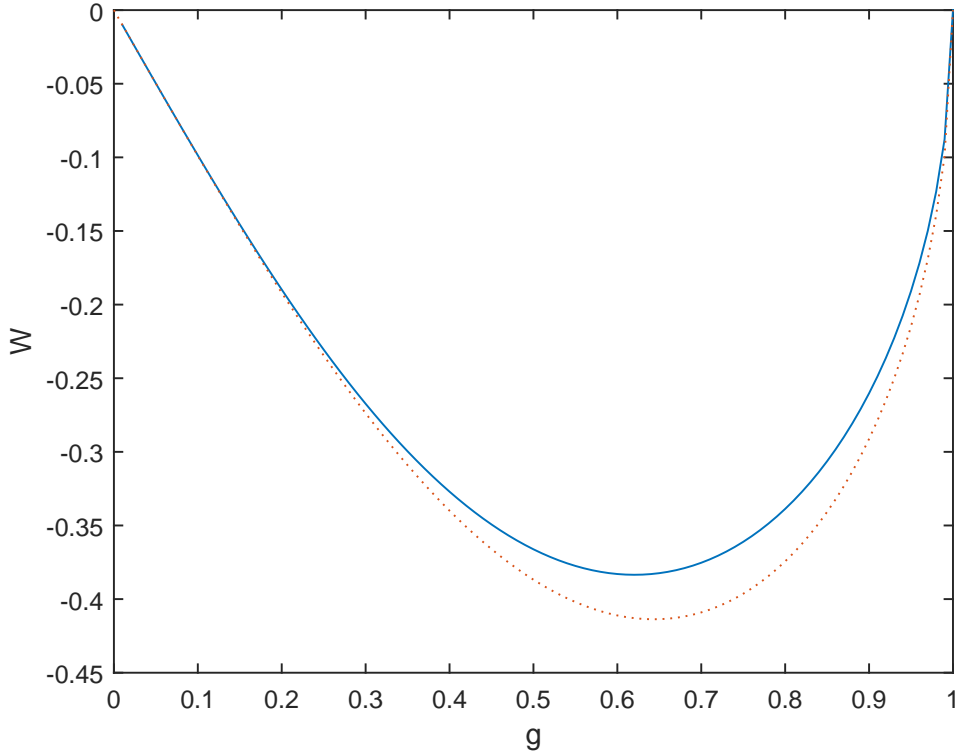

Figure 7: $W$ versus $g$ with $\mathscr{F}_2$ included (solid) compared with $W = -a$ (dotted).

### 4.7. Additional considerations

We need to examine the possible influence of the neglected terms in our calculation of $Q_A + W$. Recall that

$$g^4 \mathscr{F}_2 = WW_\sigma + Q_{AA}Q_\sigma - Q_A Q_{A\sigma}. \tag{166}$$

We earlier found that $W = -a$ under the neglect of $\mathscr{F}_2$. Now, with $\mathscr{F}_2$ retained, we are led to the following equation for $W$:

$$W = -a - \frac{\sigma^{\frac{-5\gamma}{3\gamma+1}}}{(3\gamma+1)g^4} \int_g^1 \sigma^{\frac{2\gamma-1}{3\gamma+1}} WW_g \, dg. \tag{167}$$

This is particularly simple to evaluate when $\gamma = 1/2$. Then

$$W = -\frac{4C}{k\sqrt{10}} g \sqrt{\frac{1-g^2}{1+g^2}} + \frac{2}{\sqrt{10k}} g(1-g^4)^{-1/2} W^2. \tag{168}$$

We compare the exact and approximate results in figure 7. At the nose, again for $\gamma = .5$, there is a slight error, $W \approx -1.4\sigma$ being the correct value, $W \approx -1.58\sigma$ being the approximate value when $\mathscr{F}_2$ is neglected.

Finally, we want to determine the error made in the neglect of the $A^2$ term in $\mathscr{F}_2$; see (166). That is, we need to solve

$$\mathscr{L}_2(F) = J^{-1}(Q_{AA}Q_\sigma - Q_A Q_{A\sigma}) \tag{169}$$

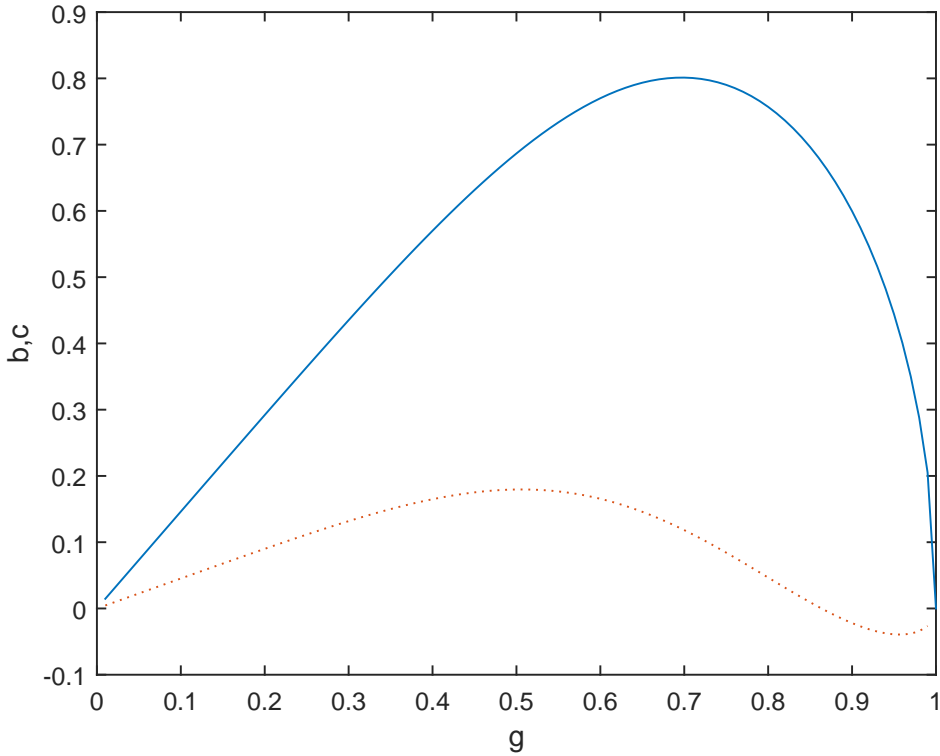

Figure 8: $b(g)$ (solid) and $c(g)$ (dotted) for $\gamma = .25$ and with $k^2 = 3$.

in the form $F = c(\sigma)(Ag^6)^2$ and, with $Q_A = b(\sigma)Ag^6$, compare $c$ with $b$ as functions of $g$. We will do this for $\gamma = .25$, under renormalization for an appropriate $k^2$. We obtain the following equation:

$$\tfrac{7}{4}\sigma(cg^{12})_\sigma + (\tfrac{9}{4} - 7\sigma g^{-1}g_\sigma)cg^{12} = -\tfrac{1}{4}g^{-4}(b^2g^{12})_\sigma. \tag{170}$$

The appropriate solution is

$$c = \frac{1}{\sqrt{7}k}g(1-g^8)^{-9/14}\int_g^1 g^{-10}(1-g^8)^{1/7}\,\frac{d(b^2g^{12})}{dg}\,dg. \tag{171}$$

We compare $b$, which is $Q_A$ without the $A^2$ correction, with the correction $c$, as functions of $g$, in figure 8, with $k = \sqrt{3}$, $C = 1.675$. The neglect of this term in $\mathscr{F}_2$ appears to be quite reasonable given the level of approximation we have been willing to tolerate in this study of the effect of axial flow.

## 5. Analysis of the local 2D problem

### 5.1. A family of generalized Sadovskii vortices

We have seen how in the absence of axial flow, the vorticity distribution in a propagating dipole must tend to the Sadovskii form, of constant values in each eddy, both in the axisymmetric case in I, and in more general settings in section 2. Recall that the Sadovskii vortex may be defined as a vorticity distribution, $\omega(x,y)$ or $\omega(r,\theta)$ in Cartesian or polar coordinates, which is constant $\pm\Omega$ in two lobes. (These local $(x,y)$ coordinates are not to be confused with those of figure 1(c).) The upper lobe is denoted $D_+$ and is bounded by the $x$-axis and a curve $C_+$, which we also express conveniently as $r = C_+(\theta)$, $0 \le \theta \le \pi$. We assume mirror symmetry of the flow about the $x$-axis with reflected region $D_-$ and curve $C_-$, setting $D = D_+ \cup D_-$ and $C = C_+ \cup C_-$. The resulting Sadovskii vortex is depicted in figure 5(a) of I and is a stationary configuration in an external flow $\mathbf{u} \sim (U,0)$ (note $U < 0$) at great distance, with lobes each having area $A_0 \simeq 37.11\,U^2\Omega^{-2}$. We recall that the process of erosion leads to the discontinuity in vorticity across $C$ (asymptotically in time), and the constancy of vorticity in each eddy arises from any smooth initial distribution following systematic erosion of outer layers.

In the presence of axial flow we still expect the process of erosion (from conservation of kinetic energy) to result in a local two-dimensional vortex with an effectively discontinuous vorticity field, zero outside a pair of lobes $D_\pm$, and we always retain the mirror symmetry. Although now there is no reason for the vorticity field to be constant in each lobe, the vorticity must remain constant on streamlines, as the vortex dipole is at leading order a solution to the two-dimensional Euler equations, only evolving (and eroding) on a longer time-scale. This motivates consideration of what we term generalized Sadovskii vortices, in which the vorticity in each lobe is $\pm\Omega + \varepsilon f(\psi)$, where $\Omega$ and $\varepsilon$ are constants, and $f(\psi)$ is some given odd function. Ideally this function would result from the full axial flow problem. This is at present too ambitious: we take $f(\psi) = \psi$ and our goal is to determine the family of vortices as prototypes to estimate approximately the effect of axial flow on the full hairpin configuration. Other approaches to computation of non-Sadovskii propagating dipoles may be found in [2] and [29].

To summarize, the family of generalized Sadovskii vortices we consider is given by

$$\omega(x,y) = \nabla^2 \psi = \begin{cases} \Omega\,\mathrm{sign}(y) + \varepsilon\psi(x,y), & (x,y) \in D, \\ 0, & (x,y) \notin D, \end{cases} \qquad (172)$$

$$\psi(x,y) = 0, \qquad (x,y) \in C, \qquad (173)$$

$$\psi(x,y) \sim -Uy, \qquad r \to \infty, \qquad (174)$$

and each member is given by parameters $\{\Omega, \varepsilon, U\}$. We take $U < 0$ without loss of generality and note that $\omega > 0$, $\psi < 0$ in the upper lobe $D_+$; thus taking $\varepsilon < 0$ strengthens the vorticity in each lobe, while $\varepsilon > 0$ suppresses it. Alternatively, we can write the problem in an integral form

$$\psi(x,y) = \int_D K(x-x', y-y')\,(\Omega\,\mathrm{sign}(y') + \varepsilon\,\psi(x',y'))\,dx'\,dy' - Uy, \qquad (175)$$

$$\psi(x,y) = 0, \qquad (x,y) \in C, \qquad (176)$$

with $K(x,y) = (4\pi)^{-1}\log(x^2+y^2)$ inverting the Laplacian in the usual way.

Note that while $\varepsilon = 0$ gives the Sadovskii vortex, taking $\Omega = 0$, $\varepsilon = -k^2 < 0$ instead recovers the classic Lamb–Chaplygin dipole, see [3] §7.3.

$$\psi = -CJ_1(kr)\sin\theta, \quad J_1(ka) = 0, \quad ka \simeq 3.83, \quad U = -\tfrac{1}{2}CkJ_0(ka), \qquad (177)$$

having area per lobe $A_0 = \frac{1}{2}\pi a^2 \simeq -23.06\varepsilon^{-1}$. Members of the family specified in (172–176) are not independent as we can scale space and time variables, so for any solution $\{\Omega, \varepsilon, U, \omega(x,y), \psi(x,y), C(\theta), A_0\}$, further solutions are given by

$$\{\Omega^*, \varepsilon^*, U^*, \omega^*, \psi^*, C^*(\theta), A_0^*\} = \{T^{-1}\Omega, L^{-2}\varepsilon, LT^{-1}U, T^{-1}\omega, L^2T^{-1}\psi, LC(\theta), L^2A_0\} \tag{178}$$

with any constants $T$ and $L$. This scaling freedom makes it clear that we seek a branch of solutions which for $\Omega, \varepsilon > 0$ interpolates between Sadovskii and Lamb–Chaplygin vortices. However we need to take $\Omega > 0$ and $\varepsilon < 0$ to discuss the effects of axial flow, and so follow the branch going from the Sadovskii vortex in the 'opposite direction' to the Lamb–Chaplygin vortex.

### 5.2. Numerics

We approach the problem numerically and outline the fixed-point iteration we will use. We suppose that at a given step we have a streamfunction $\psi^{(n)}(x,y)$ defined on an $(x,y)$-grid and know numerically the bounding contour $C_+^{(n)}(\theta_j)$ (on which $\psi^{(n)} = 0$) for a number of values $\theta_j$. We can then substitute these on the RHS of (175) and evaluate the LHS as the new approximation $\psi^{(n+1)}(x,y)$. For the RHS of (175), the piecewise constant component is taken as a contour integral over $C_+^{(n)}$ (e.g., see equation (4.24) of I), while the integral involving $\varepsilon\psi(x',y')$ is evaluated by summing over grid points $(x',y')$. To evaluate the new contour $C_+^{(n+1)}(\theta_j)$ using (176) we seek numerically a root $r$ of the equation $\psi^{(n+1)}(r\cos\theta_j, r\sin\theta_j) = 0$ for each $\theta_j$, again using (175) to evaluate $\psi^{(n+1)}$ as needed. The iteration starts either with $\{C_+^{(0)}, \psi^{(0)}\}$ as the Sadovskii vortex, or with a previously obtained solution on the same branch.

The iteration as outlined above does not work, in that iterates tend to grow or shrink in area, and rapidly drift away from the solution branch. To stabilize this, the following approach was adopted. A 'target area' was specified, say $A_{\text{target}}$, at the outset: after each iteration the area of the vortex was measured using $C_+^{(n)}$ and then the contour and the streamfunction were rescaled using a scaling factor $S_{\text{rescale}}$ to regain the target area. Iterating over $n$ as described in the previous paragraph, we obtain convergence of the stream function $\psi_+$ and bounding contour $C_+$ for this target area, and obtain a function $S_{\text{rescale}}(A_{\text{target}})$. We then use a root-finding routine to achieve $\log S_{\text{rescale}} = 0$ by varying the target area, and so obtain the solution we require, with $A_0 = A_{\text{target}}$ when $S_{\text{rescale}} = 1$. This method of stabilising the iterative scheme worked out as far as $\varepsilon = 0.4$, after which point the value of $S_{\text{rescale}}$ failed to converge. It is likely that this represents a failure of the numerical method rather than a bifurcation of the branch, and we plan to attempt other means to follow the branch further.

We give some of the results of the calculations in table 1. The first line, with parameters $\{U, \omega, \varepsilon\} = \{1, 0, -1\}$ recovers the Lamb-Chaplygin vortex with approximately the correct area (given in brackets), and the third line is the Sadovskii vortex. The second line is an intermediate case, with $\varepsilon < 0$, and the stream lines are depicted in figure 9(a), while figure 10(a) shows a scatter plot of vorticity $\omega$ against stream function $\psi$ to verify that the code has converted to the correct relationship in (172). The rest of the table gives results where we fix $U = -1$ and $\Omega = 1$ and increase $\varepsilon$ up to 0.4, where we cease to be able to follow the branch. Some stream function and scatter plots are shown in figures 9 and 10. The stream function does not change markedly in structure, but becomes slowly more elongated as $\varepsilon$ is increased.

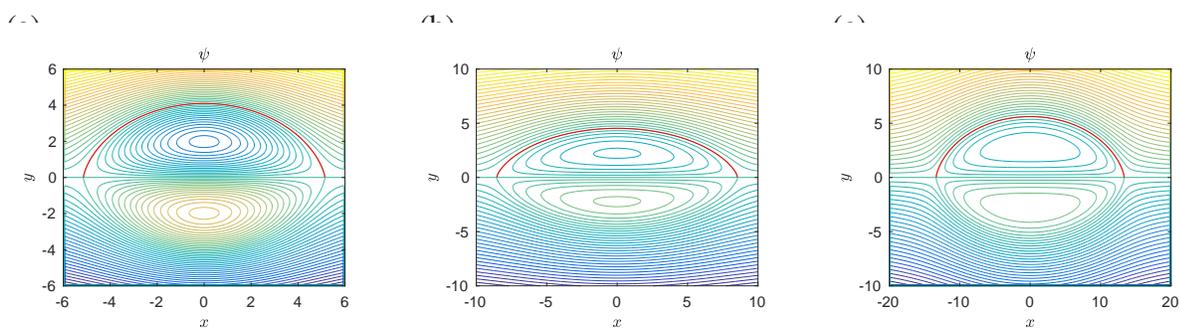

Figure 9: Stream function contour plots for $U = -1$, (a) $\Omega = 0.5$, $\varepsilon = -0.5$, (b) $\Omega = 1$, $\varepsilon = 0.2$, and (c) $\Omega = 1$, $\varepsilon = 0.35$. The curve $C_+$ is overlaid in red.

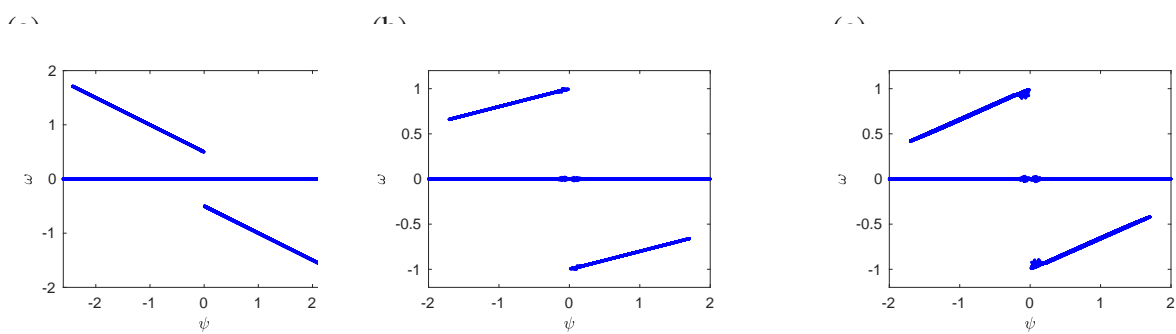

Figure 10: Vorticity versus stream function scatter plots for $U = -1$, (a) $\Omega = 0.5$, $\varepsilon = -0.5$, (b) $\Omega = 1$, $\varepsilon = 0.2$, and (c) $\Omega = 1$, $\varepsilon = 0.40$.

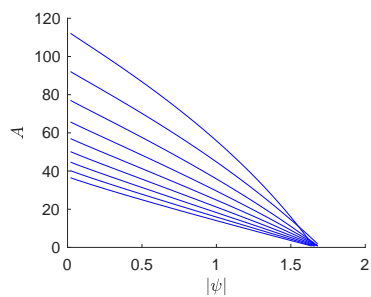

Figure 11: Area $A(\psi)$ plotted against $|\psi|$ for $U = -1$, $\Omega = 1$ and $\varepsilon = 0, 0.05, 0.1, \ldots, 0.40$ (reading up the curves).

Figure 11 shows a plot of area $A(\psi)$ inside contours of constant $\psi$ with $A(0) = A_0$ and $A(-\psi_{\max}) = 0$ defining the minimum value $-\psi_{\max}$ of the stream function for the upper lobe, say. We observe that, as a reasonable approximation over the range of $\varepsilon$ explored, the relationship between $A$ and $\psi$ is linear. This means that in each member of the family the vorticity is (approximately) a linear function of the area, going from $\omega = 1$ at the outer boundary $\psi = 0$, $A(\psi) = A_0$, to $\omega_{\min} = 1 - \varepsilon \psi_{\max}$ at $A(-\psi_{\max}) = 0$ in the center of the closed streamlines. These considerations allow us to map the family we have obtained to the family given in (164), namely

$$\omega^* = \sqrt{37.11}[1 + c^* A^* (\psi^*)/A_0^*], \quad A_0^* = 1 \tag{179}$$

in which vorticity goes from $1 + c^*$ at $\psi^* = 0$, $A^* = A_0^*$, to 1 at $\psi^* = -\psi_{\max}^*$, $A^* = 0$. In comparing the vortices from the numerically obtained one and that in (179) we can immediately identify $1 + c^* = \omega_{\min}^{-1}$ to give the correct range of vorticity. Then, to make the scaling precise we have to multiply lengths by $L = A_0^{-1/2}$ and times by $T = \omega_{\min}/\sqrt{37.11}$ to give the new velocity $U^* = LU/T$; see (178). The resulting quantities $c^*$ and $U^*$ are given in the table. Note that with the scaling, as we increase $\varepsilon$ the area $A_0^*$ is being kept constant and the vorticity is fixed in the center of the streamlines but increasing on the periphery of the lobe, at $\psi^* = 0$; with this goes an increase in the vortex velocity $U^*$.

| $-U$ | $\Omega$ | $\varepsilon$ | $A_0$ | $\omega_{\min}$ | $c^*$ | $-U^*$ |
|---|---|---|---|---|---|---|
| 1 | 0 | $-1$ | 23.10 (23.06) | 2.882 | — | — |
| 1 | 0.5 | $-0.5$ | 31.65 | 1.711 | — | — |
| 1 | 1 | 0 | 37.09 (37.11) | 1 | 0 | 1 |
| 1 | 1 | 0.05 | 40.82 | 0.916 | 0.092 | 1.041 |
| 1 | 1 | 0.1 | 45.33 | 0.831 | 0.203 | 1.089 |
| 1 | 1 | 0.15 | 50.86 | 0.745 | 0.342 | 1.147 |
| 1 | 1 | 0.2 | 57.76 | 0.660 | 0.515 | 1.214 |
| 1 | 1 | 0.25 | 66.51 | 0.575 | 0.739 | 1.299 |
| 1 | 1 | 0.3 | 77.91 | 0.493 | 1.028 | 1.400 |
| 1 | 1 | 0.35 | 92.82 | 0.420 | 1.381 | 1.505 |
| 1 | 1 | 0.40 | 113.2 | 0.338 | 1.959 | 1.694 |

Table 1: Results for the generalized Sadovskii vortex

## 6. Discussion

### 6.1. Consistency with non-existence criteria

Since $\gamma \geq 1/4$ our hairpin is consistent with the seminal BKM criterion for blowup [4]. Equally stringent are the geometric constraints on the direction of the vorticity vector; see [14] and [15]. A key constraint here which excludes blowup is, in the case of our hairpin, that a ball $B$ exists containing $\zeta = 0$ on the center curve such that

$$\lim_{t \to 0-} \sup_B \int^t \| \nabla \hat{\omega}(\cdot, s) \|_{L^\infty}^2 \, ds < \infty, \tag{180}$$

where $\hat{\omega} = \omega/|\omega|$. Then the integrand may be estimated as of order of $a^{-2} \sim \tau^{-6\gamma}$ and so this condition is not satisfied for our range of $\gamma$, namely $1/4 \leq \gamma \leq 1/2$. This result also holds if the direction of vorticity were to be dominated by centerline curvature, for then the integrand is of order $\kappa^2 \sim \tau^{2\gamma-2}$.

A more stringent condition was later given by [16]. Applied to our hairpin, the condition in their notation can be stated as follows. Let $A \geq \gamma$. Let $M(\tau) \sim a^{-1} = \tau^{-3\gamma}$ and let $L(\tau) \sim \tau^{B^*}$ where $1 - \gamma \leq B^* \leq B$. Then if $0 < B < 1 - A$ there can be no blowup. But then $4\gamma < 1$ which is here disallowed.

Finally, we thank a referee for drawing our attention to an important test of blow-up combining the analyticity-strip method of tracking singularities with the BKM theorem; see [6]. This technique relates the smallest scales measured as a power of $\tau$, here $\tau^{3\gamma}$, with the exponent $n$ of the Fourier energy spectrum of the hairpin $E(k) \sim Ck^{-n}$, where it is assumed $n < 6$. The test is that the inequality

$$3\gamma \geq \frac{2}{6-n} \tag{181}$$

is necessary for blow-up.

Although we do not have a complete solution for the hairpin structure, much less its Fourier transform, we can give an approximate calculation which is encouraging. We compute the Fourier transform of the hairpin at the moment $\tau = \tau_0$ that the singularity forms. We are only interested in exponents so we neglect multiplicative constants (indicated generically by $C$) and consider only one-half of the hairpin. Now from the asymptotic behaviour of $g$, see (15), it follows that

$$\lim_{\tau \to \tau_0} (a, \omega, J) \sim \left( s_0^{\frac{3\gamma}{3\gamma+1}}, s_0^{\frac{-4\gamma}{3\gamma+1}}, s_0^{\frac{-4\gamma}{3\gamma+1}} \right). \tag{182}$$

We will replace the dipole cross-section by a circle. We thus consider the volume of revolution in cylindrical coordinates, given by $0 \leq r \leq a$, $0 \leq s_0 \leq M$, and carry out the integral in $r, s$ coordinates.

The Fourier transform of vorticity is then, with $\tilde{r} = r/a$,

$$\omega^\dagger \equiv \int e^{i\mathbf{k}\cdot\mathbf{x}} \omega(\mathbf{x}) \, dV$$

$$\sim \int_0^M ds_0 \int_0^1 \tilde{r} \, d\tilde{r} \int_0^{2\pi} d\theta \left[ \text{sign}(\cos\theta) \exp\left[ ik_\perp C \cos(\theta - \theta_0) s_0^{\frac{3\gamma}{3\gamma+1}} \tilde{r} + ik_\parallel C \int s_0^{\frac{-4\gamma}{3\gamma+1}} ds_0 \right] s_0^{\frac{-2\gamma}{3\gamma+1}} \right]$$

$$\sim \int_0^M ds_0 \int_0^1 \tilde{r} \, d\tilde{r} \int_0^{2\pi} d\theta \left[ \text{sign}(\cos\theta) \exp\left[ ik_\perp C \cos(\theta - \theta_0) s_0^{\frac{3\gamma}{3\gamma+1}} \tilde{r} + ik_\parallel C s_0^{\frac{1-\gamma}{3\gamma+1}} \right] s_0^{\frac{-2\gamma}{3\gamma+1}} \right], \tag{183}$$

We now set $\gamma = 1/4$ since this allows the substitution $s_0 = k^{-7/3}S$ to exhibit a definite exponent:

$$\int e^{i\mathbf{k}\cdot\mathbf{x}} \omega(\mathbf{x}) dV \sim k^{-5/3} \int_0^{Mk^{7/3}} dS \int_0^1 \tilde{r} \, d\tilde{r} \int_0^{2\pi} d\theta$$
$$\times \left[ \text{sign}(\cos\theta) \exp\left[ ik_\perp k^{-1} C \cos(\theta - \theta_0) S^{3/7} \tilde{r} + ik_\parallel k^{-1} C S^{3/7} \right] S^{-2/7} \right]. \tag{184}$$

It then follows, averaging over the Fourier sphere, that $E \sim \overline{|\omega^\dagger|^2} k^2 k^{-2} \sim C k^{-10/3}$ for large $k$. We see that (181) is satisfied with the equality, as might be expected at our lower limit $\gamma = 1/4$. If now $\gamma > 1/4$ we need to apply stationary phase to the multiple integral in $\theta$ and $s_0$. The critical point is at $\theta - \theta_0 = \pi$ (the $C$'s in the exponential are positive) and a point $s_0^* > 0$. Locally the exponential takes the form $e^{ik[A\Delta\theta^2 + B\Delta s_0^2]}$, where $A, B$ involve $k_\perp/k, k_\parallel/k$. Under stationary phase we then get $\overline{|\omega^\dagger|^2} \sim E \sim C k^{-2}$. Thus (181) becomes $3\gamma \geq 1/2$ and is strongly satisfied. Note that the discontinuity introduced by $\text{sign}(\cos\theta)$ contributes an $O(k^{-1})$ term in the $\theta$ integration, and so is negligible compared to the $O(k^{-1/2})$ contribution from stationary phase.

This is a very simplified estimate and it remains to be seen if a more accurate Fourier analysis remains consistent with (181). Note that the powers here indicate that, should the singularity actually exist, its spectrum function is quite distinct from that of fully developed turbulence, although there could be some effect through intermittency corrections. The recent work of Agafontsev, Kuznetsov, and Mailybaev [1] finds evidence of Kolmogorov scaling in pancake-shaped vortical structures. The strained spiral vortex model of Lundgren [27], on the other hand, finds a $k^{-5/3}$ spectrum in tubular vorticity regions.

We have thus not found any *a priori* condition which invalidates our model. One can hope that the deeper one delves into the dynamics of the Euler equations the less likely one is to run foul of such constraints, but in the absence of a rigorous proof and the general delicacy of the problem, a future non-blowup condition would be one way to refute our proposal.

There is always lurking in studies such as this the fact that the natural instability of Euler flows might break apart the careful construct of the theorist, especially in view of our similarity assumptions, the assumed "steady" form of our collapsing hairpin, and the asymptotic quasi-two-dimensionality imposed in our formulation. We have noted in I that the elimination of the top-down symmetry of the dipole leads to a breakup in the axisymmetric case. This particular instability is perhaps less of a concern for the hairpin since, while leading to some deformation away from the symmetric state, it may develop too slowly to have a sizable effect before the singularity occurs. Even with top-down stability imposed, the possible instability of the locally 2D dipole, and the effect on such instabilities of the rapid stretching at the nose of the hairpin have not been considered in our analysis. The stability of the eroding Sadovskii dipole which we observed in the numerical studies in I is however encouraging of stability of the hairpin.

### 6.2. The question of infinite energy

Since our hairpin extends to infinity, the total kinetic energy is formally infinite. The question then is, does this provide a non-physical source for a blowup? There is a considerable literature concerning infinite energy Euler flows which blow up in finite time, see e.g. [32], [18], [17], and more recently [28]. Since these have the property that the velocity field blows up on every point of a non-compact spatial region at the singularity time, it is clear that they are non-physical owing to the source of kinetic energy at infinity.

The situation with our hairpin is quite different, owing to the conservation of kinetic energy central to the dynamics. For example for the basic hairpin, dipole area goes like $g^{-6}$, velocity like $g$, and the center curve Jacobian like $g^4$. With the initial Sadovskii cross section distributed uniformly, the initial total kinetic energy from the nose to a Lagrangian position $\zeta_0$ is then simply a multiple of $\zeta_0$. And it remains essentially at that value up to the singularity. At the singularity, vorticity and velocity are infinite at a single point at the nose. The compactly defined initial condition, mentioned in the summary subsection below, will in fact remove the issue of infinite kinetic energy entirely, but apart from that the formation of the singularity is a local phenomenon uninfluenced by distant parts of the hairpin.

### 6.3. Numerical evidence of super-exponential vorticity growth

There is unfortunately little firm evidence for a blowup of the kind proposed here in the large body of numerical experimentation involving anti-parallel vortex tubes. There has been considerable controversy surrounding the existence or not of a finite-time singularity in $\mathbb{R}^3$, see e.g. [5], [22]. Recent studies of the statistics of higher-order norms of the vorticity [19] point strongly toward a double-exponential growth [23]. The dynamical reason for this shift in growth to super-exponential is not yet well understood, but appears to be accompanied by changes in the curvature and divergence of vortex lines [23]. One goal of the present research was to possibly identify a mechanism which suppresses finite time blowup. One of our results is the failure to have found such a mechanism within the limitations of our study.

While it is possible that a singular solution of the kind we propose may exist, it could be unstable to the loss of quasi two-dimensionality, and thus suppress a finite time blowup and realize the super-exponential growth emerging in the numerical studies.

We do want to emphasize the substantial differences between the hairpin model and the initial conditions used for the most part in these numerical studies. The initial condition of [23], for example, consists of two circular tubes of length 100 and radius $a = .75$, whose centers are a distance 1.5 from the plane of symmetry. A small deformation over a length of about 5 is introduced at the center of the tubes, which we might compare with the "nose" of the hairpin. The vorticity distribution is proportional to $a^2/(r^2 + a^2)^2$. Although the resulting deformation near the nose does involve the formation of a "tail", there is no indication of the propagation of a coherent dipole. It is of course possible that given enough time a hairpin-like structure would emerge, but then it is not clear that the super-exponential growth would be maintained. In short the hairpin structure, should it be observed in a numerical experiment, could require a vastly different initial condition, as discussed below in out summary section.

### 6.4. Higher-order norms of vorticity in the hairpin

We have noted previously the role that higher-order norms of vorticity have played in evaluating the possibility of blow-up of solutions of the Euler equations. Following Gibbon [19] we consider the integrals

$$\mathscr{D}_m(t) = \left[ V_0^{-1} \left( \int_V \left| \frac{\omega}{\omega_0} \right|^{2m} dV \right) \right]^{\frac{1}{4m-3}}, \quad m = 1, 2, \ldots, \tag{185}$$

where $V_0$ is a reference volume. Gibbon then derives a differential inequality for the Euler equations:

$$\frac{d\mathscr{D}_m}{dt} \leq C_m \left(\frac{\mathscr{D}_{m+1}}{\mathscr{D}_m}\right)^{\frac{4m+1}{2}} \mathscr{D}_m^3. \tag{186}$$

We shall consider these integrals taken over the hairpin, assuming the geometry of the basic hairpin. We then have the asymptotic decay $g \sim O(\sigma^{\frac{-\gamma}{3\gamma+1}})$ for large $\sigma$. If now $m \geq 2$ and with $V_0 = a_0 U_0 \tau_0$ we find, assuming constant vorticity eddies, that

$$\mathscr{D}_m = 2\left(\frac{\tau}{\tau_0}\right)^{\frac{5\gamma+1-8\gamma m}{4m-3}} \left(\int_0^\infty g^{8m-2} d\sigma\right)^{\frac{1}{4m-3}}. \tag{187}$$

For $m \geq 2$ we see that the integral converges since $\gamma \geq 1/4$ and the integrand decays as $\sigma^{\frac{-(8m-2)\gamma}{3\gamma+1}}$. We then find that the LHS of (186) behaves as $(\tau/\tau_0)^{\frac{5\gamma+1-8m\gamma}{4m-3}-1}$ while the RHS goes as $(\tau/\tau_0)^{\frac{5\gamma+1-8m\gamma}{4m-3}-4\gamma}$. Thus (186) is for $m \geq 2$ satisfied maximally with the equality when $\gamma = 1/4$ and is otherwise satisfied strongly with an inequality.

If $m = 1$ we must deal with the divergence and so can only compute $\mathscr{D}_1$ for a fixed Lagrangian segment including the nose. This divergence introduces a factor $(\tau/\tau_0)^{3\gamma-1}$ which cancels the existing factor $(\tau/\tau_0)^{1-3\gamma}$ so $\mathscr{D}_1 \sim O(1)$. The right-hand side of (186) is then $O(\tau/\tau_0)^{\frac{1-11\gamma}{2}}$ so the inequality is strongly satisfied.

Of course here the growth of the moments tracks the singularity of the hairpin, so our point is only that we do not see any conflict with Gibbon's analysis. The approach is valuable for interpreting numerical approaches to a possible singularity in Euler or Navier-Stokes solutions [23, 24]. As we have previously noted there is unfortunately no indication at present that the above estimates for the hairpin are realized in these computations.

### 6.5. Absence of blowup for Navier-Stokes

Supposing now that we are dealing with a viscous fluid, it is of interest to examine the implications of the hairpin geometry. They depend crucially upon the invariants which are assumed. If local invariance of kinetic energy is dropped, we may consider the case $\beta = 2$, considered in [10], to replace the case $\beta = 4$ studied in this paper. With $\beta = 2$ the relevant substitutions are

$$U = -U_0(\tau_0/\tau)^\gamma g(\sigma), \quad \omega = \omega_0(\tau_0/\tau)^{2\gamma} g^2, \quad a = a_0(\tau/\tau_0)^\gamma g^{-1}, \quad \gamma \geq 1/2. \tag{188}$$

Comparing $\mathbf{u} \cdot \nabla \omega$ with $\nabla^2 \omega$, we see that both are $O(\tau_0/\tau)^{4\gamma}$. Thus a blowup of a Navier-Stokes hairpin is not ruled out, provided that energy conservation is not local.

On the other hand, with local invariance of kinetic energy and thus $\beta = 4$ we see that $\mathbf{u} \cdot \nabla \omega$ is of order $(\tau_0/\tau)^{8\gamma}$ while $\nabla^2 \omega$ is of order $(\tau_0/\tau)^{10\gamma}$. Thus viscous stresses would defeat any finite time blowup.

This result, that the augmented viscous stress resulting from the lateral erosion of the vortical eddies prevents the blowup, is not surprising in the present context. But it is at odds with any model of blowup which treats the viscous terms as a small correction to Euler's equations or a model thereof, see e.g. [33]. We emphasize the important role of local energy conservation in making the viscous stresses ultimately dominant and vortex reconnection inevitable.

While the issue of Navier-Stokes regularity remains open, there is the related question of solutions which in some sense maximize the growth of vorticity. Lu and Doering [26] have examined this question in one setting and it is interesting that their variational methods lead to structures resembling colliding vortex rings.

### 6.6. Continuation past $t = 0$

Let us suppose that for some smooth initial condition a hairpin singularity occurs at $t = 0$. What happens next? Although we would like to approach this problem *in the context of Euler's equations*, it is clear that, with velocity infinite at the singularity, the physical relevance of this flow model is lost. Let us therefore adopt the incompressible Navier–Stokes equations as the "real" fluid, and consider a sequence of solutions of the resulting initial value problem, corresponding to decreasing values of kinematic viscosity $\nu$, or rather increasing values of some Reynolds number $Re = U_0 a_0 / \nu$. Let us further assume that for any finite $Re$ the Navier-Stokes solution is defined globally in time. When we consider this limit for the hairpin, we find that, once $Re$ is large, the viscous terms become important when $\tau_0 / \tau = O(Re^{\frac{1}{27}})$; see the previous subsection. In the limit this interval defines a "cut" in the time line where Euler's equations do not apply. The customary view is that it is within the cut that diffusion of momentum leads to vortex reconnection and hence a change of the topology that created the singularity. On the far side of the cut, following a period of dissipation and as the flow adjusts to the reconnected configuration, a new well-behaved Euler flow emerges, at least out to the next singularity and cut. For example a typical reconnection would be between the vortex lines on each side of the nose of the hairpin, across the plane $y = 0$, allowing release of reconnected vortex tubes on either side of the nose. Since $a = O(Re^{-3/2})$ at reconnection the circulation in these tubes at the point of reconnection would be $O(Re^{-1})$.

The question we wish to raise is, in the limit of infinite $Re$ is the jump of the Euler flow across the cut uniquely defined by the initial condition? Of course in a practical sense this question is academic. Deterministic chaos renders the flow essentially random irrespective of the existence of a singularity. Nevertheless, given total control over the initial condition, we may suppose the Euler flow to be completely determined for all time in the absence of a singularity. What we are considering here is a higher level of sensitivity, wherein what transpires within the cut cannot ultimately (in the inviscid limit) be controlled by the initial condition. We do not know the answer to our question, but consider it sufficiently important to mention here. A negative answer would imply that the Euler flow is unpredictable after a finite time.

### 6.7. Summary

In this paper there are three main results, all of which bear on the general question of how fast vorticity can self-stretch. First, we propose the "steady" similitude for a finite time singularity of an Euler flow, consisting of the seven equations (119)–(124) for the eleven variables $Q, \Gamma, A_\psi, H, w, W, \omega, U, \kappa, J, A_0$, plus the Poisson equation $\nabla^2 \psi = \omega$ and the associated equation $A_\psi H_A = \omega$, the determining relation for $U$, and the specification of local kinetic energy conservation. The three-dimensional plus time Euler problem is thus replaced by a system of equations of at most second order. But the main advantage of our system is that the need for computation in three dimensions at arbitrarily small scales has been eliminated.

Second, we propose that our basic hairpin, having at each cross-section a Sadovskii dipole, offers a promising initial condition for a direct numerical simulation of the primitive equations. The parameter $\gamma$ of the hairpin satisfies $1/4 \leq \gamma \leq 1/2$ but is otherwise unrestricted, so it is tempting to choose $\theta|_{g=0}$, see (20), to be $\pi/4$, allowing a patching together (with suitable smoothing) of segments of four center curves to form a rounded square, so that the resulting symmetry might be exploited. This would set $\gamma \approx .36$ and blowup of vorticity like $\tau^{-1.44}$. The dipole edge could be smoothed to make the initial condition $C^\infty$. Perhaps a variational approach for Euler such as that used in [26] for Navier-Stokes could produce a useful finite initial structure related to the hairpin.

Third, we have made a preliminary investigation of the axial flow developed in the hairpin, which is inconclusive with regard to the suppression of a finite time singularity. Although we have found an extension of the Sadovskii dipole to a family closer to what is required by our hairpin, it remains unclear if there exists a propagating dipole compatible with the steady similitude of the hairpin. We put this forward as the most likely way our model would fail to realize finite time blowup.

We thank Eric Siggia for helpful comments on this work. We also thank the referees for their insightful critiques and further references. This paper is cordially dedicated to Keith Moffatt on the occasion of his 80th birthday.